\documentclass[aps,prc,reprint,superscriptaddress]{revtex4-1}
 \usepackage{dcolumn}
\usepackage{amsmath,amssymb,amsthm,paralist}
\usepackage{graphicx}
\usepackage{appendix}
\usepackage{subfigure}
\usepackage{longtable}

\usepackage{amsmath}
\usepackage{amsfonts}
\usepackage{hyperref}
\usepackage{url}
\usepackage{verbatim}
\usepackage{xcolor}

\begin{document}


\title{Elastic scattering of \texorpdfstring{$^3$}{}He + \texorpdfstring{$^4$}{}He with SONIK}

\author{S. N. Paneru}
\email[]{sp266413@ohio.edu}

\affiliation{Department of Physics \& Astronomy, Ohio University,
Athens, Ohio 45701, USA}
\author{C.~R. Brune}
\affiliation{Department of Physics \& Astronomy, Ohio University,
Athens, Ohio 45701, USA}
\author{D. Connolly}
\affiliation{TRIUMF,
  Vancouver, British Columbia, Canada}
\author{D. Odell}
\affiliation{Department of Physics \& Astronomy, Ohio University,
Athens, Ohio 45701, USA}
\author{M. Poudel}
\affiliation{Department of Physics \& Astronomy, Ohio University,
Athens, Ohio 45701, USA}
\author{D.~R. Phillips}
\affiliation{Department of Physics \& Astronomy, Ohio University,
Athens, Ohio 45701, USA}
\author{J. Karpesky}
\affiliation{Department of Physics, Colorado School of Mines,
  Golden, Colorado 80401, USA}
\author{B. Davids}
\affiliation{TRIUMF, Vancouver, British Columbia, Canada}
\affiliation{Physics Department, Simon Fraser University, Burnaby, BC, Canada}
\author{C. Ruiz}
\affiliation{TRIUMF,
Vancouver, British Columbia, Canada}
\author{A. Lennarz}
\affiliation{TRIUMF,
Vancouver, British Columbia, Canada}
\author{U. Greife}
\affiliation{Department of Physics, Colorado School of Mines,
  Golden, Colorado 80401, USA}
\author{M. Alcorta}
\affiliation{TRIUMF,
  Vancouver, British Columbia, Canada}
\author{R. Giri}
\affiliation{Department of Physics \& Astronomy, Ohio University,
Athens, Ohio 45701, USA}
\author{M. Lovely}
\affiliation{Department of Physics, Colorado School of Mines,
  Golden, Colorado 80401, USA}
\author{M. Bowry}
\affiliation{TRIUMF,
  Vancouver, British Columbia, Canada}
\author{M. Delgado}
\affiliation{TRIUMF,
  Vancouver, British Columbia, Canada}
\affiliation{Instituto de Física Teórica IFT-UAM/CSIC, Madrid, Spain}
\author{N. E. Esker}
\affiliation{TRIUMF,
  Vancouver, British Columbia, Canada}
\author{A. Garnsworthy}
\affiliation{TRIUMF,
  Vancouver, British Columbia, Canada}
\author{C. Seeman}
\affiliation{TRIUMF,
  Vancouver, British Columbia, Canada}
\author{P. Machule}
\affiliation{TRIUMF,
  Vancouver, British Columbia, Canada}

\author{J. Fallis}
\affiliation{North Island College, British Columbia, Canada}
\author{A. A. Chen}
\affiliation{McMaster University, Ontario, Canada}
\author{F. Laddaran}
\affiliation{University of British Columbia, British Columbia, Canada}
\author{A. Firmino}
\affiliation{University of Alberta, Edmonton, Alberta, Canada}
\author{C. Weinerman}
\affiliation{McGill University, Montreal, Quebec, Canada}


\begin{abstract}
 Measurements of the elastic scattering cross section of $^3$He and $^4$He are important in order to improve constraints on  theoretical models of $^4\rm He( ^3\rm He,\gamma )^7\rm Be$, a key reaction in Big Bang nucleosynthesis and solar neutrino production. The astrophysical $S$-factor for this reaction is a significant source of uncertainty in the standard-solar-model prediction of the $^7$Be and $^8$B solar neutrino fluxes. The elastic scattering measurements reported in the literature do not extend to low energies and lack proper uncertainty quantification. 
 A new measurement of the $^4$He($^3$He,$^3$He)$^4$He reaction has been made at center-of-mass energies $E_{\text{c.m.}}=0.38 -3.13 $ MeV using the Scattering of Nuclei in Inverse Kinematics (SONIK) 
scattering chamber: a windowless, extended gas target surrounded by an array of 30 collimated silicon charged particle detectors situated at TRIUMF. This  is  the  first elastic  scattering  measurement of  $^3$He+$^4$He made below 500 keV and it has greater angular range and better precision than previous measurements. The elastic scattering data were analyzed using both $R$-matrix and Halo Effective Field Theory (Halo EFT) frameworks, and values of the $s$-wave scattering length and effective range were extracted.
The resulting improvement in knowledge of the $s$-wave effective-range function at low energies will reduce the overall uncertainty in $S_{34}$ at solar energies.
\end{abstract}


\maketitle

\section{INTRODUCTION}
\label{Introduction}
The reaction $^4$He($^3$He,$\gamma$)$^7$Be is of critical importance for the production of high energy neutrinos during pp-chain burning in low mass stars like our sun. The $^7$Be produced by this reaction undergoes electron capture to produce $^7$Be neutrinos in the pp-II chain. In the pp-III chain, the $^7$Be undergoes a radiative proton capture reaction to form $^8$B which subsequently $\beta^+$ decays  to produce $^8$B neutrinos. The total active flux of these $^7$Be and $^8$B neutrinos has been measured by the Borexino and SNO detectors with uncertainties of $\pm3\%$ and $\pm4\%$, respectively~\cite{Borexino,SNO}. The Super Kamiokande experiment also reported a measurement of the $^8$B neutrino flux with an uncertainty of $\pm3\%$~\cite{SK_2016}. However, the predicted $^7$Be and $^8$B neutrino fluxes from the calculations of the standard solar model (SSM) have uncertainties of $\pm$6$\%$ and $\pm$12$\%$, respectively~\cite{Vinyoles_2017}. The low energy astrophysical $S$ factor for the $^4$He($^3$He,$\gamma$)$^7$Be radiative capture reaction, $S_{34}$($E$), is respectively the first and second most uncertain nuclear input in the SSM prediction of the $^7$Be and $^8$B neutrino fluxes~\cite{Vinyoles_2017}. It must be known at or near the Gamow
peak energy of $\sim$18 keV, which is experimentally inaccessible due to Coulomb barrier suppression. The cross sections are unmeasurably
small at these energies, so available data starting around $E_{\text{c.m.}}= 100$ keV must be extrapolated to solar energies with the aid of theoretical models. \par
Several different theoretical approaches used to calculate the $^4$He($^3$He,$\gamma$)$^7$Be reaction cross section at these energies are summarized in the ``Solar Fusion II'' review~\cite{SolarFusionII}. In that work these approaches were sifted and a subset of them used to extrapolate the experimental capture data available in 2011.  The resulting recommended zero-energy astrophysical $S$-factor for the $^4$He($^3$He,$\gamma$)$^7$Be reaction is $S_{34}(0)$=0.56$\pm$0.02 (expt) $\pm$ 0.02 (theory) keV b. 
\par
The $^4$He($^3$He,$\gamma$)$^7$Be reaction also plays a key role in determining the quantity of ${}^7{\rm Li}$ produced by Big-Bang Nucleosynthesis (BBN)~\cite{Cyb16}. In this case, the important energy range is approximately $100\lesssim E_{\text{c.m.}}\lesssim 600$~keV~\cite{Nol00,Ili20}, a region where the cross section can be measured directly by experiment. Currently, the BBN predictions for the ${}^7{\rm Li}$ abundance are about a factor of three higher than observations, which is far too large of a difference to be explained by uncertainties in the $^4$He($^3$He,$\gamma$)$^7$Be reaction. Since the $^4$He($^3$He,$\gamma$)$^7$Be reaction rate at BBN temperatures can be determined directly from experimental data, i.e., without extrapolation or a detailed model, we do not focus on this application in the present work.
\par
The capture reaction $^4$He($^3$He,$\gamma$)$^7$Be proceeds dominantly through a non-resonant
direct capture mechanism into the ground and first excited state of $^7$Be. Measurements of the $^4$He($^3$He,$\gamma$)$^7$Be cross section have been made by detecting the prompt $\gamma$-rays, the $^7$Be activity, and the $^7$Be recoils, which could be broadly categorized into prompt and activation measurements. The results from these two types of measurements were not
in agreement until 1998, as summarized in Solar Fusion I~\cite{SolarFusionI}. With the advancement in
measurement techniques, the recent results from both types of measurement are remarkably
consistent, as pointed out in Ref.~\cite{deBoer14}. The cross section for $^4$He($^3$He,$\gamma$)$^7$Be  has been measured for center-of-mass energies from 90 keV to 3.2 MeV.\par
$^{4}$He+$^3$He elastic scattering is an important constraint on theoretical models and phenomenological descriptions of the $^4$He($^3$He,$\gamma$)$^7$Be
reaction: any model of the capture reaction should also be able to describe low-energy elastic scattering. In phenomenological descriptions such as $R$-matrix,  $^4$He($^3$He,$^3$He)$^4$He is an open channel and it affects the extrapolation of $^4$He($^3$He,$\gamma$)$^7$Be cross-section data to solar energies. In theoretical models the quality of scattering wave functions that are input to the calculation of the  $^4$He($^3$He,$\gamma$)$^7$Be reaction cross section can be assessed by the models' ability to describe the elastic scattering cross section.   

A comprehensive $R$-matrix analysis of $^4$He($^3$He,$\gamma$)$^7$Be reaction by deBoer \textit{et al}.~\cite{deBoer14} studied the effects of elastic
scattering data on the inferred astrophysical $S$-factor values at solar energies and reported $S_{34}$(0)=0.542 $\pm$ 0.023 keV b---a central value about 3$\%$ lower than is recommended in Solar Fusion II. The authors of that study emphasized the need for a new study of $^4$He($^3$He,$^3$He)$^4$He covering a wide angular range with detailed
uncertainty estimates. This conclusion is bolstered by recent \textit{ab initio} and Halo Effective Field Theory (EFT) calculations. In Ref.~\cite{DOHETERALY} the
no-core shell model with continuum (NCSMC)~\cite{DOHETERALY} was used to compute the  $^4$He($^3$He,$\gamma$)$^7$Be reaction {\it ab initio}. Dohet-Eraly {\it et al.} found discrepancies between the elastic scattering phase shifts they predict and experimental observations. 
Iliadis et al.~\cite{Iliadis2016} performed a global Bayesian estimate based on microscopic models and \textit{ab initio} methods. Scaling the model calculations  to fit the capture data, they reported $S_{34}$(0)=0.572 $\pm$ 0.012 keV b. This central value is 2$\%$ higher
than Solar Fusion II and 6$\%$ higher than that of de Boer {\it et al.}.
Meanwhile, two recent Halo EFT calculations that used ${}^3$He and ${}^4$He as degrees of freedom  showed the strong connection between the $s$-wave
scattering length and effective range and the shape of the capture-reaction $S$-factor at low energies~\cite{higa2018radiative, zhang_2020}. Ref.~\cite{zhang_2020} recommended $S_{34}(0)=0.577^{+0.015}_{-0.016}$ keV b based on a Bayesian analysis of
radiative capture data. 

It is thus clear that the $^4$He($^3$He,$^3$He)$^4$He elastic scattering is important both in astrophysics and
few-body nuclear theory. However, there are only a few experimental studies of this reaction at the low energies where the information is most pertinent to solar fusion. Most of the experiments reported in the literature are motivated to understand the structure
of $^7$Be and consequently are focused on high resonance energies~\cite{Phillips_1958, Tombrello_1963, Barnard_1964, Spiger_1967, Ivanovich_1967, Chuang_1971, Hardy_1972, Boykin_1972}. The only measurement extending to low energies reached $E[^{3}$He]=1.2 MeV, but lacks error estimates~\cite{Mohr_1993}.
\par
This paper describes a new measurement  at TRIUMF of the elastic scattering cross section of the $^4$He($^3$He,$^3$He)$^4$He reaction. The measurement was carried out at incident beam energies as low as  $E$[$^3$He]=0.721  MeV. The experimental method used to measure the elastic scattering is explained in Sec.~\ref{experiment}. In Sec.~\ref{analysis} we discuss the details of the data analysis and the calculation of the differential scattering cross section. In Sec.~\ref{sec.IV} the differential scattering cross sections from this measurement are compared to existing measurements from literature. In this paper the $s$-wave scattering length for the $^3$He+$^4$He system is determined using 
both a multilevel $R$-matrix approach and Halo EFT to simultaneously analyze the new elastic scattering data from this work and the data of Ref.~\cite{Barnard_1964}. These analyses are described in Secs.~\ref{Rmatrix} and \ref{eft}, respectively. The results of the analyses  are presented in Sec.~\ref{results}, which also
contains 
a comparison with previous results from the literature. We conclude in Sec.~\ref{conclusion}.

\section{EXPERIMENT}
\label{experiment}
The elastic scattering measurement of $^4$He($^3$He,$^3$He)$^4$He was performed at TRIUMF. A $^3$He beam in the 1+ charge state was produced using the TRIUMF Off-Line Ion Source (OLIS)~\cite{OLIS}. The beam was accelerated using the Isotope Separator and Accelerator-I (ISAC-I) facility and delivered to the the Scattering of Nuclei in Inverse Kinematics (SONIK)~\cite{Devin} apparatus with an intensity of about 10$^{12}$ $\text{s}^{-1}$. SONIK was filled with $^4$He gas maintained at a typical pressure of 5 Torr.
\par
\begin{figure}[htp]
\includegraphics[width=1.0\columnwidth,angle=0]{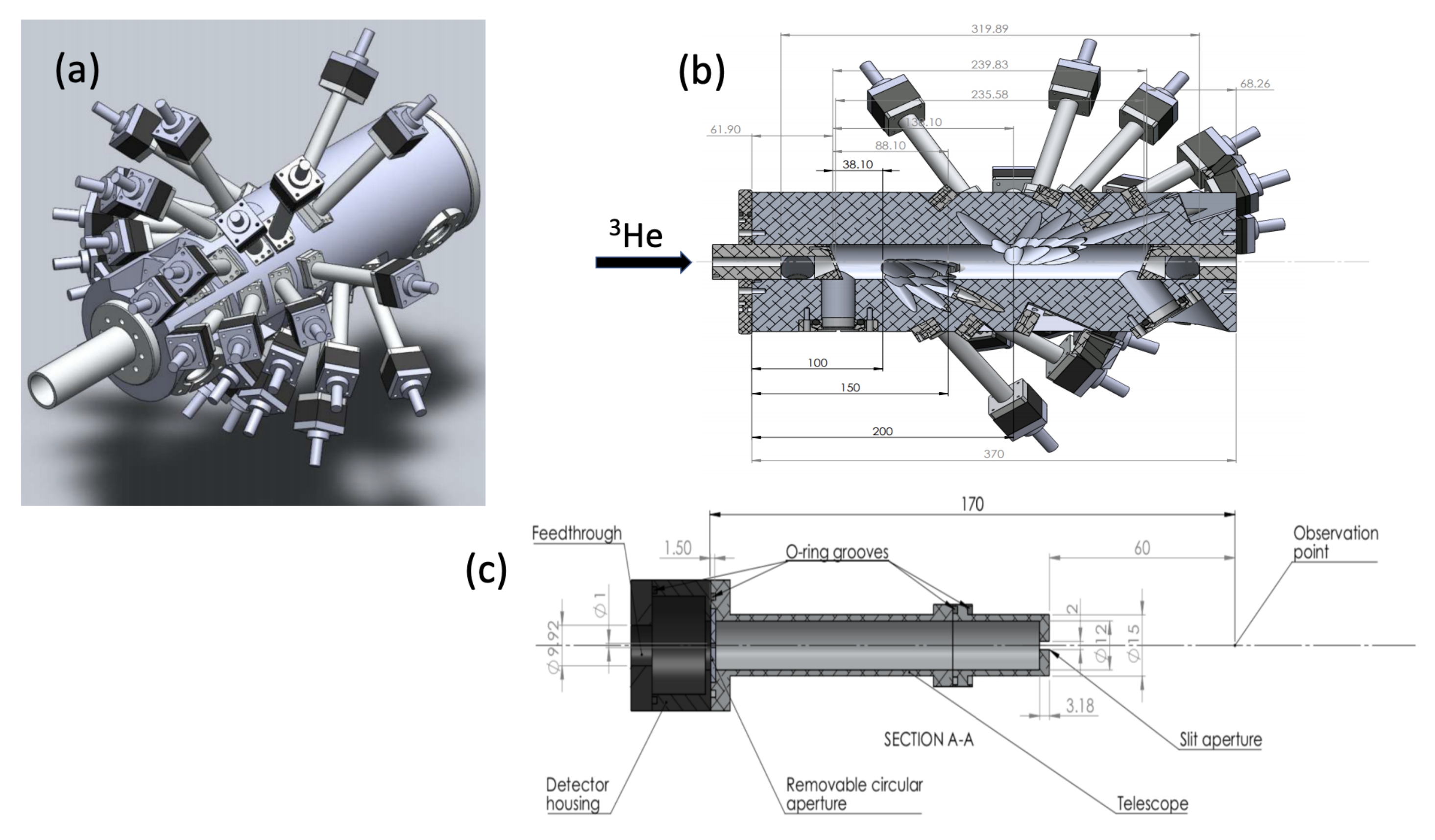}
\caption{(Color online) a) 3-D model of SONIK. b) SONIK design details. The $^3$He beam traverses the $^4$He gas target from left to right in the figure. c) A detector telescope assembly. The dimensions in the figures are in mm.}
\label{SONIK_scheme}
\end{figure}

SONIK is a windowless, extended gas
target surrounded by an array of 30 collimated silicon charged particle detectors. The chamber was commissioned with two separate measurements, $^4$He($^3$He,$^3$He)$^4$He elastic scattering and $^7$Li($p$, $p$)$^7$Li
elastic scattering. The charged particle detectors are mounted in an assembly referred to as the detector telescope hereafter. The design details of SONIK and the detector telescopes are shown in Fig.~\ref{SONIK_scheme}. 
Each detector telescope is collimated by a 2.0-mm-wide rectangular slit aperture at
the telescope’s interface with the gas volume and a 1.0-mm-diameter circular aperture in
front of the silicon charged particle detectors. The two apertures
are separated by a distance of 11.0 cm. The distance from the front aperture to the
observation point on the beam axis in the gas target is 6.0 cm. The detectors are placed at a distance of 17.0~cm from the center of the beamline, measured along the axis of the telescope, and at observation angles ranging from 22.5$^\circ$ to 135$^\circ$ in the laboratory frame. The beam delivered to SONIK enters the windowless gas target through a 6-mm diameter aperture and exits through an 8-mm-diameter aperture; they are separated by 23.98 cm. A constant pressure is achieved along the extended gas target by using the Detector of Recoils and Gammas of Nuclear Reactions (DRAGON)
differential pumping system. Helium gas was cleaned by continuous
recirculation through a LN$_2$ cooled zeolite trap.  The detector
telescopes are arranged such that they observe three different points, termed interaction
regions, in the gas target along the beam direction ($z$-direction). Since each interaction
region has a different $z$-coordinate, the bombarding energy and therefore the scattering
energy varies slightly by interaction region. This arrangement of three interaction regions
is highly beneficial for inverse kinematics experiments with radioactive beams, where
the beam time is limited and there are narrow resonances to be studied. Additional details of the experimental setup are given in Refs.~\cite{Devin, SNPaneru}.
\par
\begin{figure}[t]
\includegraphics[width=1.0\columnwidth,angle=0]{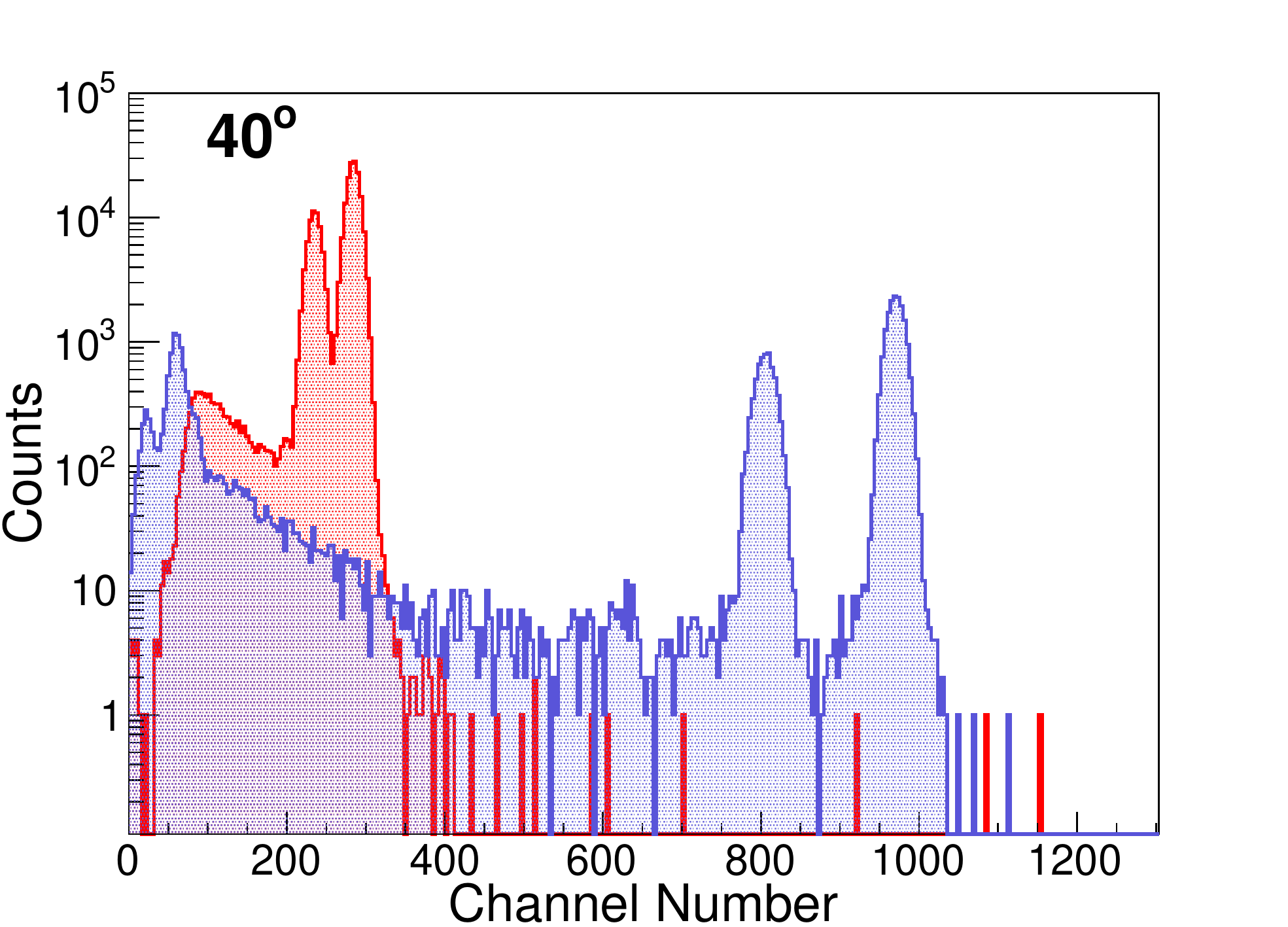}
\caption{Typical spectrum from the experiment. The red and blue histograms represent the spectra obtained at $\theta_{\text{lab}}=40^{\circ}$ for $E$[$^3$He]=1.303 MeV and 3.608 MeV, respectively. The $^3$He and $^4$He peaks are resolved at the higher incident energy but not at the lower beam energy. }
\label{raw_spec}
\end{figure}

$^4$He($^3$He,$^3$He)$^4$He elastic scattering was measured at 9 energies corresponding to
 $^{3}$He beam energies of $E[^{3}$He]=~ 0.721, 0.878, 1.303, 1.767, 2.145, 2.633, 3.608, 4.347, and 5.490 MeV. This is the first ever measurement made below $E_{\text{c.m.}}$=~0.50 MeV of $^4$He($^3$He,$^3$He)$^4$He elastic scattering. Since the projectile and target masses are comparable, we observed both the recoils and ejectiles from the elastic scattering in our detectors. A typical raw spectrum for two incident beam energies from the experiment is shown in Fig.~\ref{raw_spec}.
 The $^3$He and $^4$He peaks are well resolved at high incident beam energies. We aimed for 1000 counts in the 90$^\circ$ detector for each incident beam energy before changing to the next energy. For low incident beam energies, we couldn't observe the $^3$He ejectiles at 90$^\circ$, so the next detector to observe was then at 75$^\circ$. The 120$^\circ$ and 135$^\circ$ spectra were not used for the analysis because of their limited utility.\par
 
\section{DATA ANALYSIS}
\label{analysis}
The differential elastic scattering cross section in the laboratory frame of reference at bombarding energy $E_{0}$ and scattering angle $\theta_{0}$ 
 is given by:
\begin{equation}\label{cross_section_calculator}
    \frac{d \sigma}{d \Omega}(E_{0},\theta_{0})=\frac{N_{\rm det} \sin{\theta _0}}{n N_{\rm inc} G \epsilon},
\end{equation}
where $N_{\text{det}}$ is the number of detected particles, $n$ is the target density, $N_{\text{inc}}$ is the number of incident beam particles, $G$ is the $G$ factor~\cite{Silverstein} discussed in Section~\ref{G-factor}, and the quantity $\epsilon$ is the beam transmission through an empty gas target.
\subsection{Beam Energy Determination}
The beam energies were measured using the DRAGON facility according to the procedure given in Ref.~\cite{HUTCHEON}. The $^3$He beam in charge state $q$ was centered on a 2-mm slit downstream of
DRAGON's first magnetic dipole MD1, and the measured field
value $B$ was converted to energy per nucleon $E/A$ by using the relation
\begin{equation}\label{Energy}
\frac{E}{A}=C_{\text{mag}}\left(\frac{qB}{A}\right)^2-\frac{1}{2\mu c^2}\left(\frac{E}{A}\right)^2
\end{equation}
where $E$ is the kinetic energy of $^3$He in MeV, $C_{\text{mag}}$=48.15(7) MeVT$^{-2}$~\cite{HUTCHEON}, $A=3.016
$ is the mass of $^3$He in atomic mass units, and $\mu$ is one atomic mass unit in $\text{MeV}/c^2$.\par

\begin{figure}[htbp]
    \centering
    \includegraphics[width=1.0\columnwidth,angle=0]{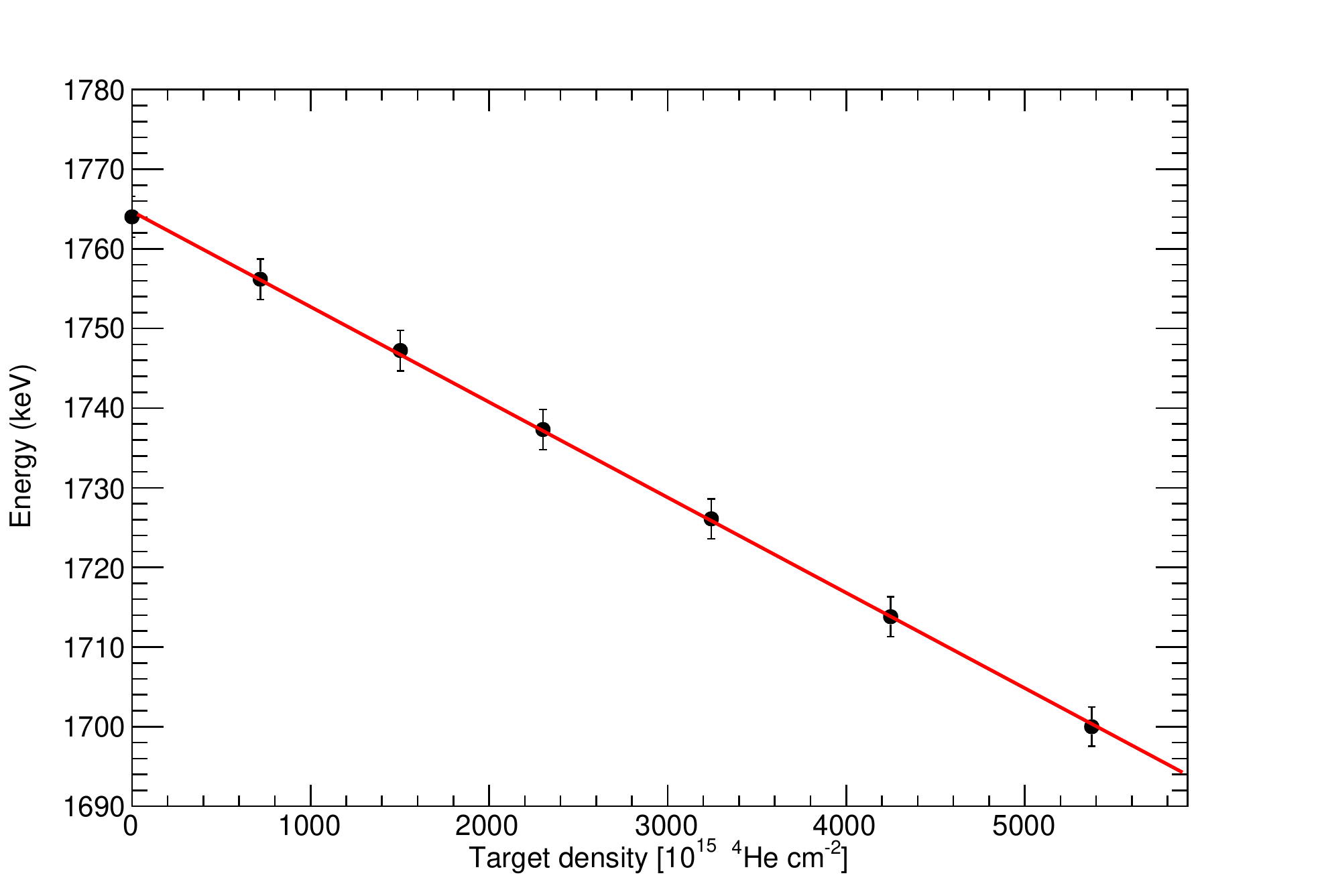}
    \caption{Determination of stopping power for $E$[$^3$He]=1.767 MeV. The error bars are entirely due to the systematic error associated with the constant $C_{\text{mag}}$ in Eq.~(\ref{Energy}).}
    \label{stopping_power}
\end{figure}

With DRAGON, we are able to make direct stopping power measurements. The beam
passing through the gas target at varying pressures is deflected by the first bending magnet (MD1)
downstream of the target and centered onto the charge slit. The magnetic field strength required
to transmit the beam through the charge slit is measured. The stopping power measurements for
SONIK were performed at $E$[$^3$He]=1.767 MeV with pinhole apertures (1.5 mm diameter)
at the beam entrance and beam exit positions as well as with the standard (6 and 8mm) apertures in order to measure the effective length of the target. The target areal density is determined from the target pressure using the ideal gas law.  The beam energy is plotted as a function of target density which yields the linear relationship shown in Fig.~\ref{stopping_power}. The slope of this line is the stopping power for $^3$He in $^4$He gas. At $E$[$^3$He]=1.767 MeV the stopping power obtained via this approach is 11.97$\pm$0.53 eV/(10$^{15}$atoms/cm$^2$). Meanwhile, that obtained from a Stopping and Range of Ions in Matter (SRIM)~\cite{SRIM} calculation is 11.08  eV/(10$^{15}$atoms/cm$^2$). The central value of the experimentally measured stopping power differs from the SRIM prediction by $8.0 \%$. The measured stopping power value is consistent within 1-$\sigma$ error bars if the uncertainty in the SRIM predictions of 4.3$\%$ ~\cite{SRIM_uncertainty} is taken into consideration. The length of the gas target with standard apertures (i.e. beam entrance and beam exit apertures with diameters of 6 mm and 8 mm, respectively) is termed the effective length in our experiment. The effective length differs from the physical length as the gas in the differential pumping system diffuses outwards, thereby increasing the length of the gas target with which the beam interacts.  With the assumption of linear energy loss, the effective target length can be determined via
\begin{equation}
l_{\text{eff}}=\frac{\Delta E}{n \mathcal{S}},
\end{equation}
where $\Delta E$ is the energy loss in the target, $\mathcal{S}$ is the stopping power of $^3$He in $^4$He, and $n$ is the target number density. The effective length for the gas target was calculated to be $l_{\text{eff}}=24.61\pm1.09$ cm.

The stopping power measurement was only performed for $E$[$^3$He]=1.767~MeV, where an 8\% difference from the SRIM calculation was found. Assuming the same difference for other incident energies, the SRIM-calculated stopping powers were increased by 8$\%$ to obtain the stopping power of $^3$He in $^4$He gas. With the assumption of linear energy loss, the effective beam energy $E_i$ at each interaction region $i$ in the gas target is calculated by
\begin{equation}
    E_{i}=E_{\text{in}}-\mathcal{S} t_{i},~~~ \text{with}~~~~~ i=1,2,3
\end{equation}
where $E_{\text{in}}$ is the incident beam energy, $\mathcal{S}$ is the scaled stopping power and $t_{i}$ is the target areal density for interaction region $i$. The errors on the incident energy, effective
length, measured stopping power, temperature, and pressure are propagated to calculate
the error on the energy at each interaction region. The error was calculated to be 6.6$\%$ of
the energy loss, which amounts to a maximum of 3 keV in $E [^{3}\text{He}]$ for the lowest $^3$He incident beam
energy at interaction region III.
\subsection{Transmission}
The incident beam currents are measured upstream and downstream of the SONIK chamber using Faraday cups FC4 and FC1, respectively. The ratio of FC4 to FC1 gives us the measure of beam transmission through the target. With no gas in the target, the ratio FC4/FC1 should be ideally 1, which is not the case practically. The exercise of measuring the beam transmission was done at the start of each energy change but was not measured for each individual run. It is believed to be a constant factor for all runs for a given incident beam energy. The empty gas beam transmission for $E$[$^3$He]=5.490, 3.608, 2.633, and 1.303 MeV were measured again before making an energy change. The mean of the difference of the transmission measured at the start and at the end of run before the energy change for these energies was found to be 3$\%$. Therefore, for all energies, an additional 3$\%$ uncertainty was added in quadrature with the counting errors in the empty gas beam transmission measurement.
\subsection{Beam Normalization}
\label{sec:beam_normalization}
The FC4 Faraday cup reading is used to determine the number of incident beam particles $N_{\text{inc}}$. The target density of $^4$He, $n$, is determined from the pressure and temperature of the gas target. The product of the number of incident beam particles, target density, and the empty gas beam transmission $\epsilon$ gives the normalization for each energy measurement. The beam normalization error for each energy is the error associated with the normalization and is a common mode error for all data points for a given energy. There was a change in incident
beam intensity in between the runs while acquiring data for$E[^{3}$He]= 2.633 MeV resulting in a different normalization. So, there are two set of runs of data for $E[^{3}$He]= 2.633 MeV. The common mode errors for the different energies lie between 3.7 and 9.6$\%$ and are listed in Table~\ref{chi_square}. 
\subsection{Yield Measurement}
The individual peak yields from the raw spectrum were extracted after background subtraction. The yields for both the $^3$He and $^4$He peaks were calculated whenever possible. The low-energy feature seen in the typical spectrum from the experiment shown in Figure~\ref{raw_spec} is background due to the contribution of detector noise and electrons produced from in-target scattering which tail off gradually with increasing energy. Beside these sources, a contribution to the background also arises from energy-degraded ions from beam scattering off the aperture edges upstream of the SONIK interaction regions, the energy degraded scattered particles from the edges of the Si detector collimators, and particles backscattered out of the Si detector before depositing all of their energy.  The last two sources of background were included in the {\sc{Geant4}} simulation, but the effects were too small to explain the observed tails on the $^3$He and $^4$He peaks. When the peaks are fully resolved, polynomials were used to characterize the background, which then were integrated analytically to estimate the background contribution to the peak yields. When the peaks overlap, two Gaussian functions with a common exponential or polynomial background function are used to fit the spectrum and extract the peak yields. The random errors for each peak yield were determined considering the errors due to the choice of fit parameters, the use of a Gaussian function to fit the peak, and calculating yields for each one hour spectrum versus the yields for the summed spectrum for a given beam energy. The random background error estimates from each of these three components were added in quadrature to calculate the total random background error. The random background error was estimated for each peak for all energies and was added in quadrature with the statistical error to obtain the reported point-to-point error in the differential scattering cross section.
\subsection{G factor}
\label{G-factor}
For charged particle scattering experiments with a gas target and collimated detector telescopes, the relationship between the number of detected particles, $N_{\text{det}}$, and the differential scattering cross section, $\frac{\text{d}\sigma}{\text{d}\Omega}$, is often expressed as
\begin{equation}\label{G-factor1}
    N_{\text{det}}=\frac{n~N_{\text{inc}}~G~}{\sin{\theta _0}}\frac{d \sigma}{d \Omega}(E_0,\theta_0),
\end{equation}
where $n$ is the areal density of target nuclei, $N_{\text{inc}}$ is the number of incident beam particles,  $E_0$ is the beam energy,  $\theta_0$ is the central angle subtended by the detector system, and the acceptance of the detector is given by the $G$ factor instead of the usual solid angle.The inclusion of the sine term makes $G$ independent of $\theta_0$ at  leading order.
\par
\begin{figure}[htbp]
    \centering
    \includegraphics[width=1.0\columnwidth,angle=0]{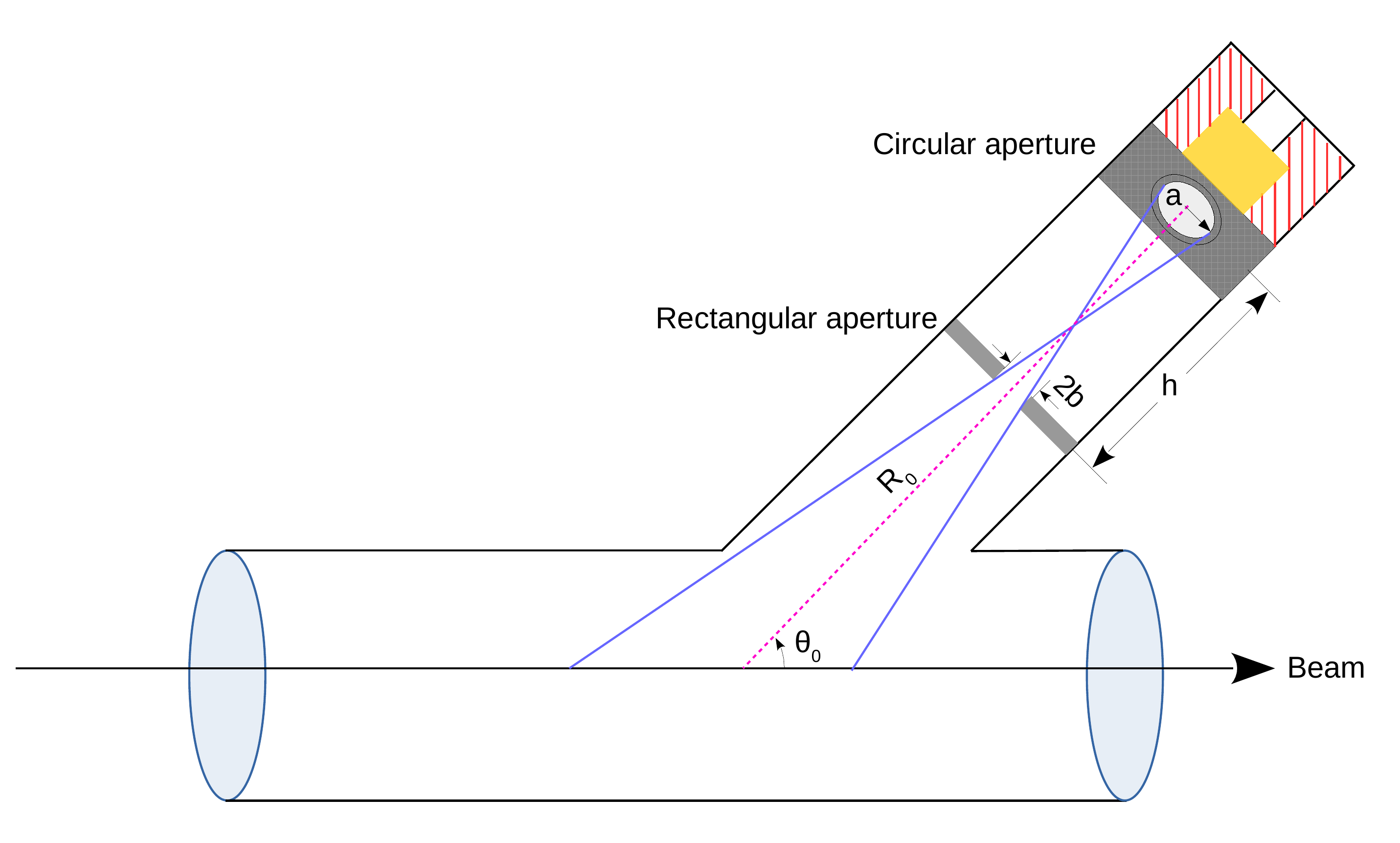}
    \caption{A detector telescope of SONIK.}
    \label{schematic}
\end{figure}

For the SONIK chamber, we have the front aperture as a vertical slit of width 2$b$, where the slit is perpendicular to the plane defined by the beam axis and the center line of the detector system. The rear aperture is circular with radius $a$. The distance between the apertures is $h$ and the distance from the central interaction point on the $z$-axis (beam axis) to the rear aperture is $R_0$. A schematic diagram of SONIK doubly-collimated apertures is shown in Fig.~\ref{schematic}.
Assuming $a,b\ll R_{0}$ and  $a,b\ll h$, the analytical expression for the $G$ factor for a collimated detector like that of SONIK is given by Silverstein~\cite{Silverstein} as
\begin{equation}\label{Silverstein}
    G=G_{00}(1+\Delta_0),
   \end{equation}
 where
\begin{equation}\label{leading_order}
       G_{00}=\frac{2\pi a^2b} {R _0 h},
\end{equation}
and
\begin{equation}
       \Delta_{0}=\frac{a^2 \cos^2{\theta_0}} {4R^2 _0 \sin^2 \theta_0}-\frac{b^2}{2h^2}-\frac{3a^2}{8}\left[\frac{1}{h^2}+\frac{1}{R^2_0}\right].
\end{equation}
The quantity $\Delta_0$ incorporates second order corrections in the ratio of $a$ and $b$ to either  $R_{0}$ or $h$. 
Eq.~(\ref{Silverstein}) includes the geometrical effects on the acceptance of the detector system, however the acceptance also depends on slit edge scattering, multiple scattering, beam size, beam divergence, etc. A charged particle traversing the gas target undergoes numerous small angle collisions (screened elastic scattering) with the gaseous atoms termed ``multiple scattering". In our experiment, the effect of multiple scattering could be visualized in two processes. First, the incident beam particles undergo multiple scattering, effectively increasing the beam diameter. Second, a charged particle from the elastic scattering undergoes multiple scattering before it is detected in the Si detector, changing the acceptance of the detector. The elastic scattering of $^3$He+$^4$He was measured as low as $E$[$^3$He]=721 keV in this experiment, at which the effect of multiple scattering is expected to be the greatest. The multiple scattering of the incident beam particle would affect the overall acceptance. To account for these effects, particularly multiple scattering, a Monte Carlo simulation was developed in the {\sc{Geant4}}~\cite{GEANT4} framework to calculate the $G$ factor for SONIK. The $G$ factor calculated from the simulation also includes the effects of the energy and angular spreads of the beam.
\par
\begin{figure}[tbp]
\includegraphics[width=1.0\columnwidth,angle=0]{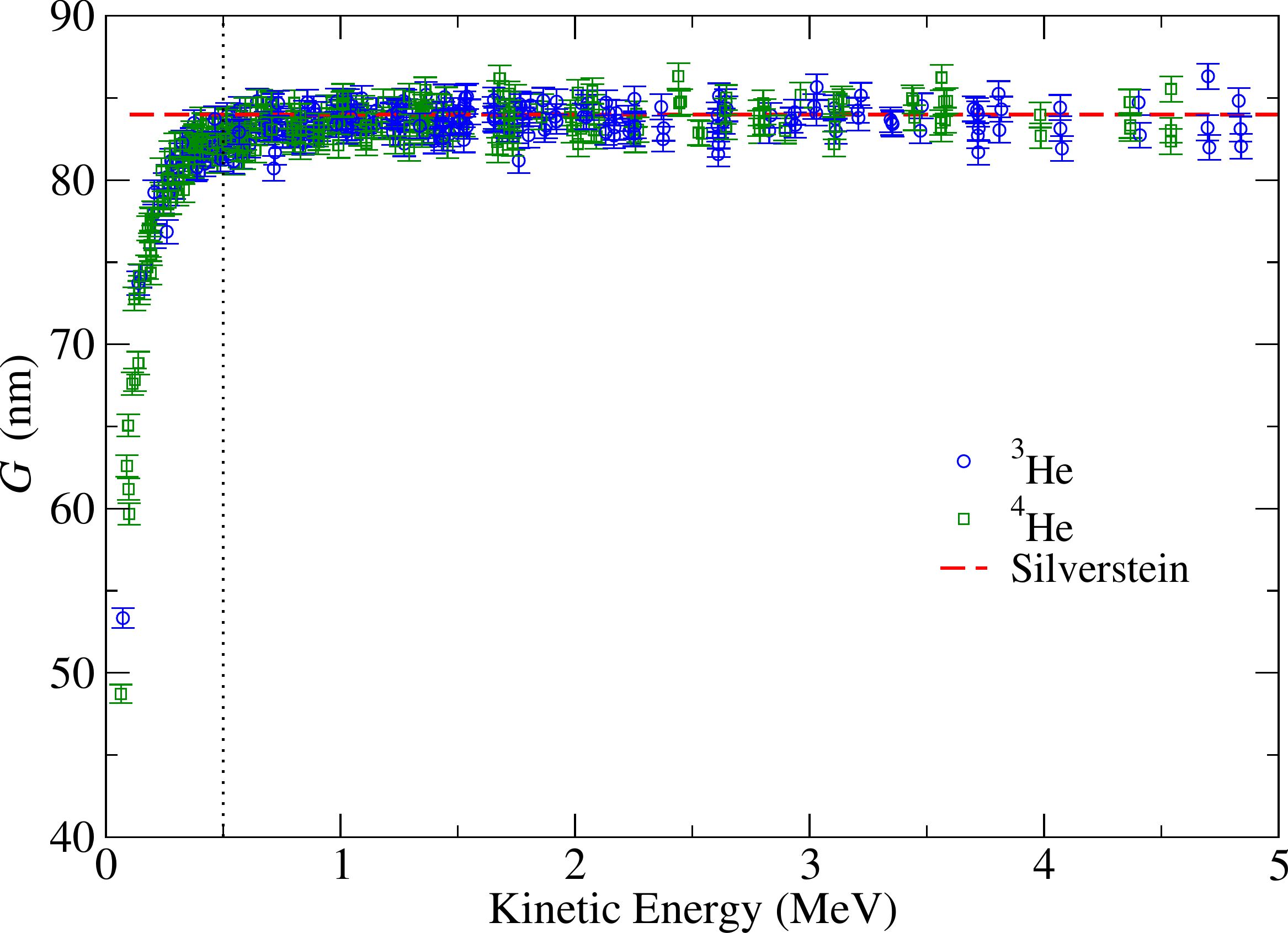}
\caption{The $G$ factor as a function of kinetic energy of the scattered particle. The blue circles and green squares are the computed $G$ factor from the {\sc{Geant4}} simulation for $^3$He and $^4$He, respectively. The red-dashed line is the $G$ factor calculated using Eq.~(\ref{Silverstein}). The dotted line is explained in the text.}
\label{G_vs_E}
\end{figure}
\begin{figure}[tbp]
\includegraphics[width=1.0\columnwidth,angle=0]{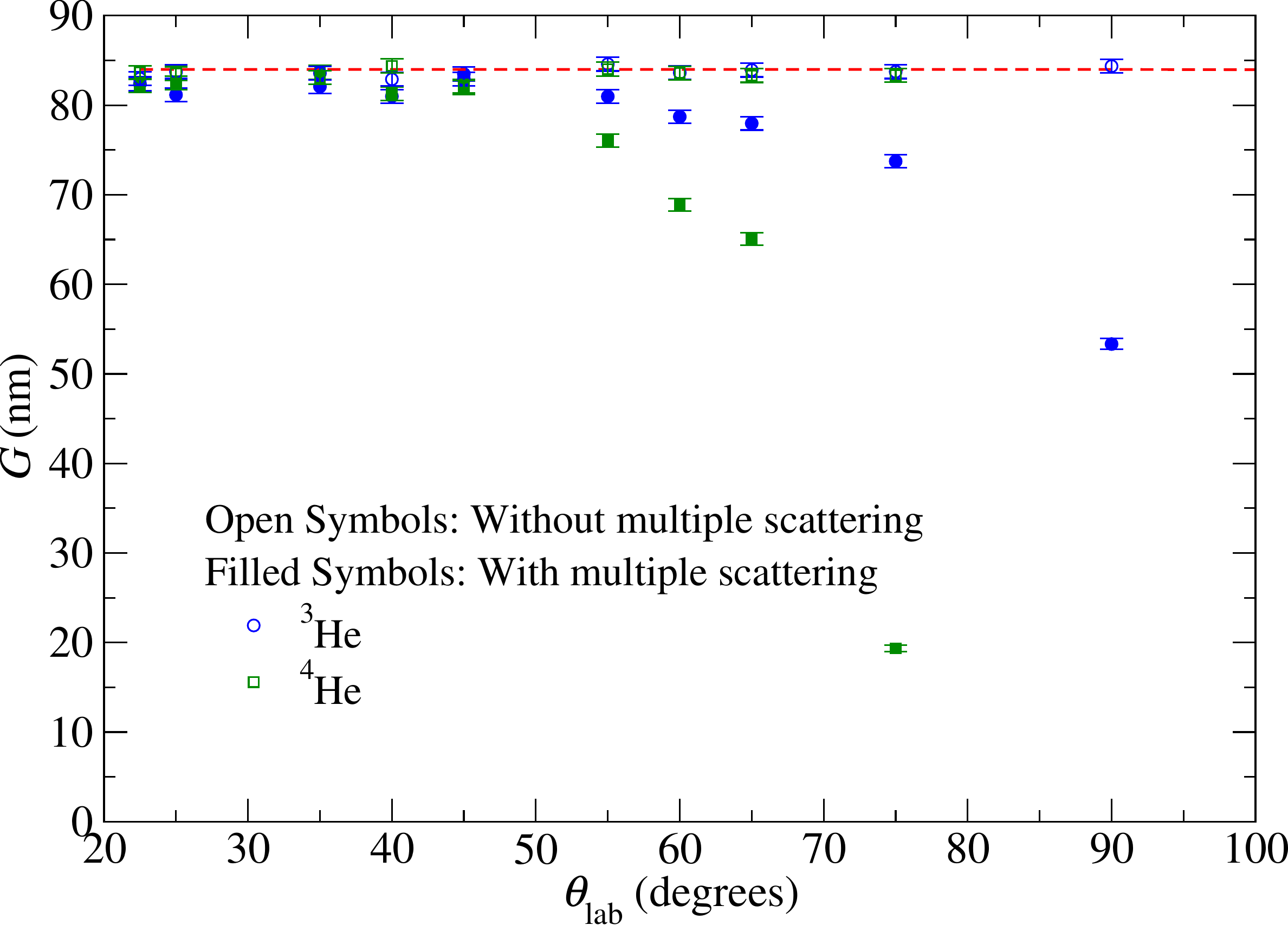}
\caption{The $G$ factor as a function of scattering angle in laboratory frame of reference for $E$[$^3$He]=721~keV. The filled (open) points represent the $G$ factor obtained from a {\sc{Geant4}} simulation with (without) introducing the multiple scattering model in the simulation. The red-dashed line is the  $G$ factor calculated using Eq.~(\ref{Silverstein}).}
\label{msc_vs_nomsc}
\end{figure}

The {\sc{Geant4}} simulation was performed in two steps. First, the $^3$He beam particles were introduced along the $z$-direction through
the $^4$He gas target kept at a temperature $T= 30^{\circ}$C and pressure $P=5$ Torr. The trajectory of each beam particle is stored. We used the G4Urban Msc-Model~\cite{G4urbanmsc} to simulate the multiple scattering effects of the $^3$He particles in the $^4$He gas target. The $^3$He beam introduced in the $^4$He gas target loses energy as it traverses the target. The stopping power of $^3$He in the $^4$He gas target calculated from the simulation was in good agreement with SRIM calculations~\cite{SRIM}.
Second, the scattered events were generated using the information from the stored tracks and the scattered particles detected by the Si detectors.
The tracks are chosen randomly from the stored beam particle trajectories. The scattered 
particle properties such as position and energy are extracted from the chosen track. For a line beam without multiple scattering and assuming an energy- and angle-independent differential cross section, i.e.,
$\frac{d\sigma}{d\Omega}=\frac{\sigma}{4\pi}$, the product $n N_\text{inc} \sigma$ in Eq.~(\ref{G-factor1}) gives the number of reactions per unit length along
the beam path, which provides a link between the simulation and the $G$ factor. A Monte
Carlo simulation is implemented by generating events randomly from a uniform distribution along the length $\Delta z=z_{\text{max}}-z_{\text{min}}$, along the beam axis and randomly from a uniform distribution into a solid angle $\Delta \Omega=(\rm cos~ \theta_{\text{min}}-\rm cos~\theta_{\text{max}})(\phi_{\text{max}}-\phi_{\text{min}})$. The parameters $z_{\text{min}}$, $z_{\text{max}}$, $\theta_{\text{min}}$, $\theta_{\text{max}}$, $\phi_{\text{min}}$, and $\phi_{\text{max}}$ are determined using the geometry as in Ref.~\cite{Silverstein}. The energy of the generated scattering event $E$ is randomized
within the energy limit, $E_{\text{min}}$ and $E_{\text{max}}$, using the inverse transform method assuming the probability of scattering is inversely proportional to the square of the energy. The energy limits $E_{\text{min}}$ and $E_{\text{max}}$ were determined from the stored tracks and correspond to the energy of the beam particle at $z_{\text{min}}$ and  $z_{\text{max}}$, respectively. For a given energy, the corresponding position $\vec{r}$ of
the scattering event was obtained from the stored tracks.
The generated event with co-ordinates ($\vec{r}$,$\theta$,$\phi$,$E$) was accepted based on the acceptance-rejection method. Let $N_{\text{ev}}$ be the number of events generated along $\Delta z$ into the solid angle $\Delta\Omega$ and $N_{\text{det}}$ be the number of events detected in the Si detector in the {\sc{Geant4}} simulation, given as
\begin{equation}\label{G-factor2}
    N_{\text{ev}}=nN_{inc}\frac{d \sigma}{d \Omega}(E_0,\theta_0)\Delta z\Delta \Omega.
\end{equation}
From Eq.~(\ref{G-factor1}) and Eq.~(\ref{G-factor2}), the $G$ factor can be computed as
\begin{equation}\label{G_factor_simulation}
    G=\frac{N_{\text{det}}}{N_{\text{ev}}}\rm sin\theta_{0}\Delta z\Delta \Omega.
\end{equation}
The simulation was run for $N_{\text{ev}}$=10$^6$ events. The plot of the $G$ factor as a function of energy of the scattered particle is shown in Fig.~\ref{G_vs_E}. The $G$ factors for both $^3$He and $^4$He particles show the same behaviour as a function of energy.
 The calculated $G$ factor from the simulation is in good agreement with the values obtained with the analytic expression, Eq.~(\ref{Silverstein}), starting around particle energies above 1 MeV. However, if we go further down in particle energy, the differences between the $G$ factor from the simulation and the analytic expression increase and become significant for energies below 500 keV as represented by the dotted line in Fig.~\ref{G_vs_E}. The lowest particle energy for which the $G$ factor from the simulation is used is 0.24 MeV, for which the $G$ factor is 80.6 nm.  Note that our results are consistent with Eq.~(\ref{Silverstein}) if we do not introduce the multiple scattering effects in our simulation. These results with multiple scattering switched off were a benchmark for our simluation, and they are shown in Fig.~\ref{msc_vs_nomsc}. 
\par
The choice of multiple scattering model is one of the sources of systematic uncertainty
in the $G$ factor derived from the Monte Carlo simulation. We used two models, namely the
G4Urban Msc-model~\cite{G4urbanmsc} and the Wentzel-VI Msc-model~\cite{Ivanchenko_msc_2010} to simulate the multiple scattering. The systematic uncertainty in the $G$ factor due to the choice of multiple scattering model in this work was approximately $\pm 1\%$. 
\par
It is often the case that the geometrical components of SONIK, such as the apertures, will have some errors associated with their dimensions when made in the machine shop.
Measurements of the rectangular apertures and circular apertures in each of the detector telescopes were made with an optical comparator or shadow graph.  The rectangular aperture dimensions, $b$, are on average 0.1$\%$ larger than the specified value of 1.0 mm and have a standard deviation of 0.4$\%$ around the mean. The circular aperture radii, $a$, are on average 2.2$\%$ larger than the specified value of 0.5 mm and have a standard deviation of 1.6$\%$ about the mean. The error on $R_{0}$ and $h$ is negligible compared to the error on $a$. In Sec.~\ref{bayesian_analysis} the variation in the aperture dimensions of each detector is modeled using a detector-dependent normalization, to which we assign
a Gaussian prior with a mean of 0.96 and standard deviation of 0.032 (cf. Eq.~(\ref{leading_order})). This implies an additional 27 normalization factors, $c_{j}$, corresponding to the 27 detectors that were actually used to collect the data.


\subsection{Error Budget}
This section describes the systematic error for all measurements made in the experiment independent of angle and energy. In our experiment, the beam is tuned through the SONIK chamber using the CCD camera and the beam profile monitors. However, the beam position might change during the run period, which changes the acceptance of the detector or $G$ factor. The effect of a change in beam position on the calculated differential elastic scattering cross section was estimated to be $\pm$1$\%$, as explained in detail in Ref.~\cite{SNPaneru}.
The systematic uncertainties due to various other factors are presented in Table~\ref{systematic_uncertainty_1}.
The individual systematic uncertainties are added in quadrature to report the total systematic uncertainty for all measurements from this experiment. The total systematic uncertainty for all measurements is estimated to be within 2.0$\%$.
\begin{table}[tbp]
\renewcommand{\arraystretch}{1.5}
\caption{Estimation of systematic uncertainties.}
\label{systematic_uncertainty_1}
\begin{tabular}{lc}\hline\hline
Source of Error & Value  \\\hline
Target pressure and temperature & 1$\%$ \\
Beam intensity & 1$\%$ \\
Beam position & 1$\%$ \\
Model Uncertainty in \sc{Geant4} & 1$\%$\\\hline
Total & 2.0$\%$\\\hline\hline
\end{tabular}
\end{table}
\par
The differential scattering cross sections from this work are shown in Figs.~\ref{Fits_SONIK_1}, \ref{Fits_SONIK_2}, and~\ref{Fits_SONIK_3}. The red circles and purple squares represent the differential scattering cross sections calculated using $^3$He and $^4$He ejectiles, respectively.

\section{Comparison with previous data}\label{sec.IV}

\begin{figure}[htp]
    \includegraphics[width=1.0\columnwidth,angle=0]{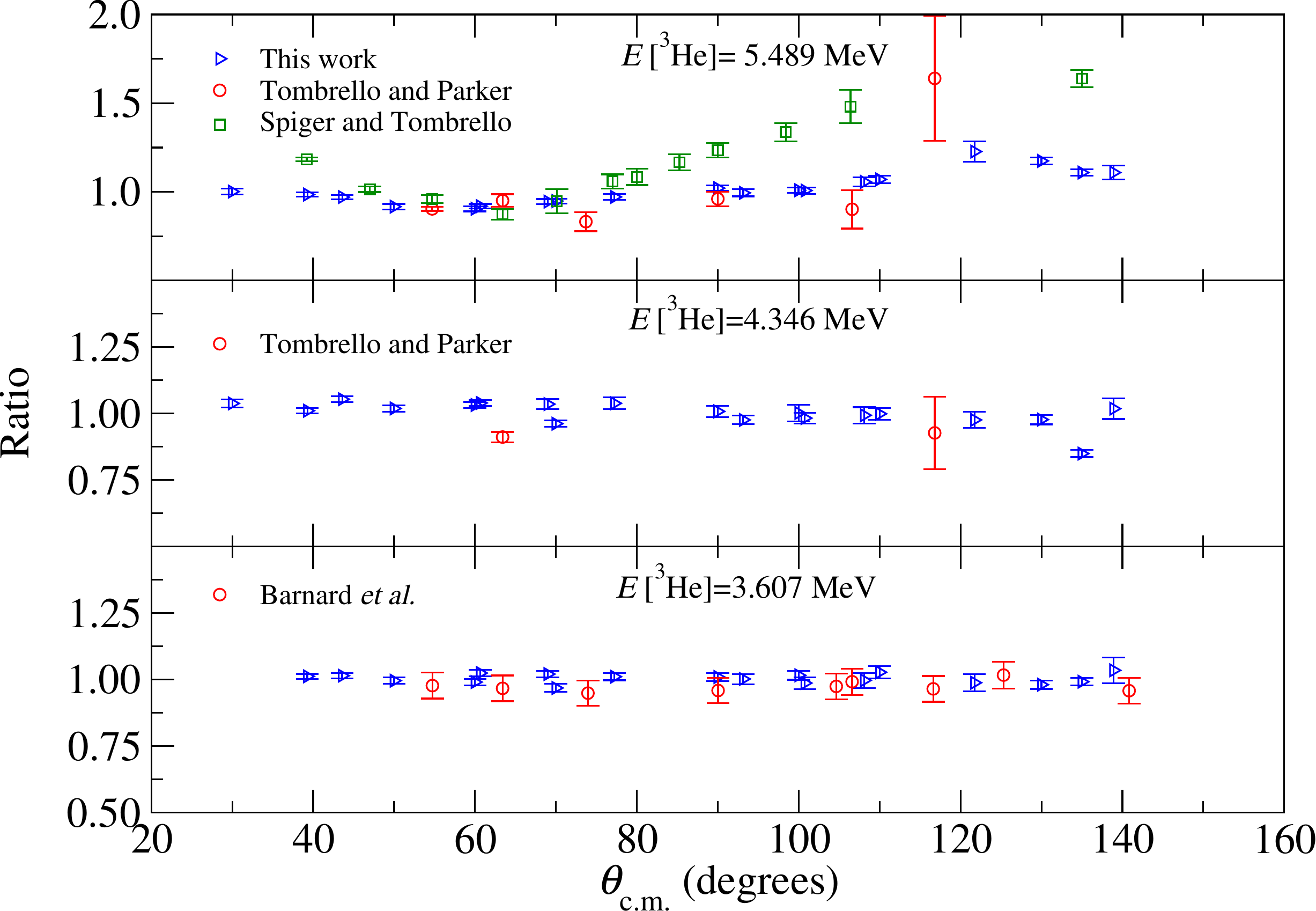}
    \caption{The ratio of the experimental  differential scattering cross section to the cross section calculated using $R$-matrix parameters from Ref.~\cite{deBoer14} at beam energies of 5.490 MeV, 4.347 MeV, and 3.608 MeV. The interaction region III measurements of this work, represented by the blue points, are, in general, consistent with previous determinations but are more precise.  Only interaction region III data are shown for these comparisons. }
    \label{comparison_1}
\end{figure}

\begin{figure}[htp]
    \includegraphics[width=1.0\columnwidth,angle=0]{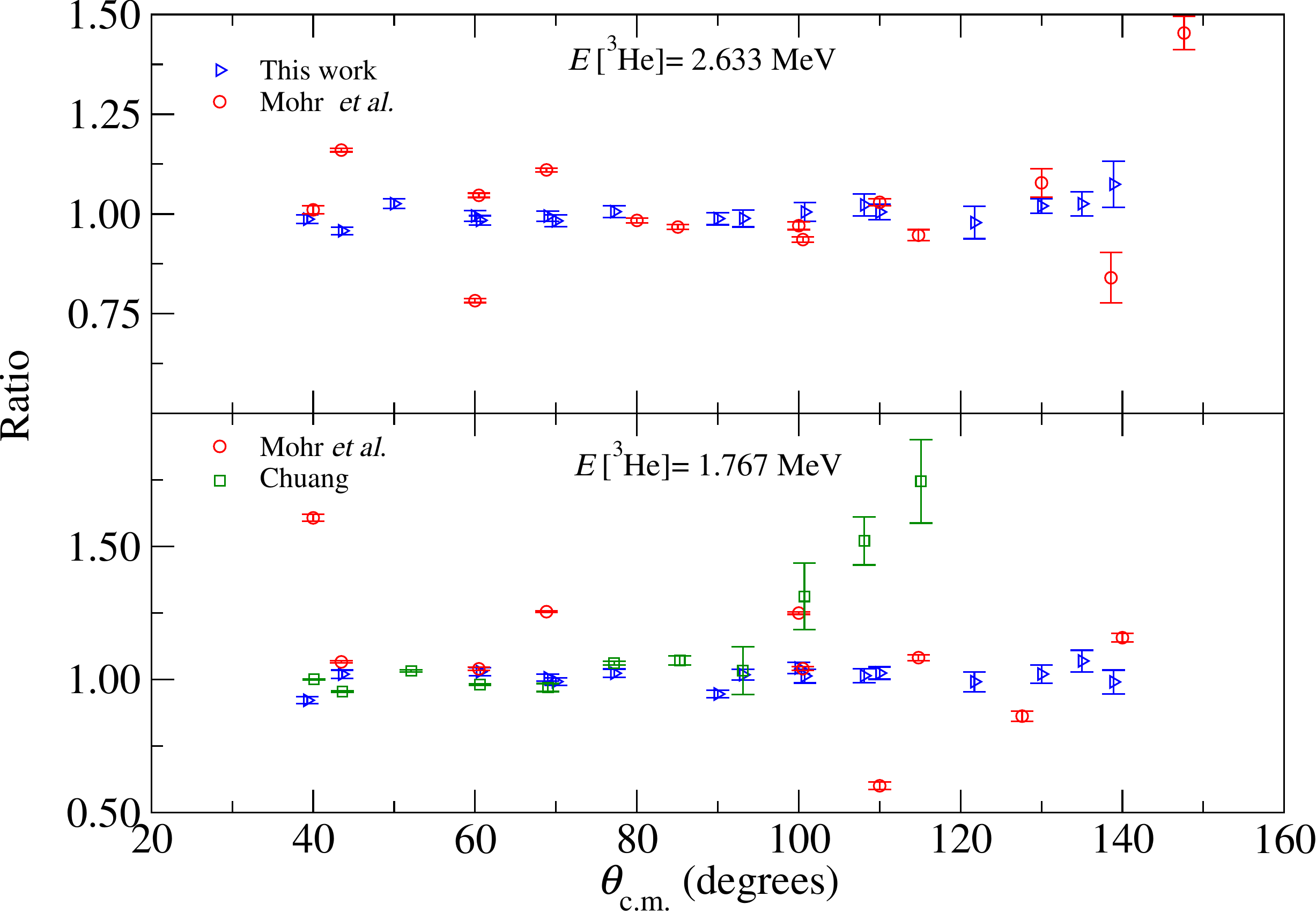}
    \caption{The ratio of the experimental  differential scattering cross section to the cross section calculated using $R$-matrix parameters from Ref.~\cite{deBoer14} at beam energies of 2.633 MeV and 1.767 MeV. The discrepancies between the interaction region III measurements of this work, represented by the blue points, and previous data are discussed in the text. }
    \label{comparison_2}
\end{figure}

To compare our result with previous measurements of elastic ${}^3$He-${}^4$He scattering whose energy range overlaps that of our experiment we use the ratio of the experimental differential scattering cross section to the cross section calculated using the $R$-matrix parameters determined in Ref.~\cite{deBoer14}. That ratio is plotted in Fig.~\ref{comparison_1} (beam energies of 5.490 MeV, 4.347 MeV, and 3.608 MeV) and Fig.~\ref{comparison_2} (beam energies of 2.633 and 1.767 MeV). Overall the results from this work are consistent with previous determinations but have better precision. 
The data from Spiger and Tombrello~\cite{Spiger_1967} shown in the top panel of Fig.~\ref{comparison_1} are from a measurement at $E$[$^3$He]=5.438 MeV, a slightly different energy compared to our measurement. The lowest panel in Fig.~\ref{comparison_1} shows that the data from this work are in good agreement with those of Barnard \textit{et al}.~\cite{Barnard_1964}.
Turning to Fig.~\ref{comparison_2},
the Mohr \textit{et al}.~\cite{Mohr_1993} measurements at $E$[$^3$He]=2.6 MeV and $E$[$^3$He]=1.7 MeV show jumps in between the data points, whereas our result corresponds to a smooth angular distribution.
The lower panel of Fig.~\ref{comparison_2}
shows that the present result is in fair agreement with data from  Chuang~\cite{Chuang_1971} at forward angles, although three backward-angle data points from Ref.~\cite{Chuang_1971} disagree. 

The Spiger and Tombrello~\cite{Spiger_1967},  Tombrello and Parker~\cite{Tombrello_1963}, and Chuang~\cite{Chuang_1971} measurements were made using a gas cell with foil entrance and exit windows. The energy loss corrections for the charged particles at the entrance and exit windows of the gas cell introduce additional systematic uncertainties in the energy determinations for these older measurements. The use of a windowless gas target in our work avoids the need for the energy loss corrections. This could account for  discrepancies seen in the cross section at backward angles between our results and the past measurements. The Mohr \textit{et al}.~\cite{Mohr_1993} measurement was performed with a jet gas target and the scattered particles were detected using 10 surface barrier detectors placed at fixed positions. $^{20}$Ne was mixed with the $^{4}$He gas in the target for the normalization purposes. This measurement does not specify its systematic uncertainty, so the differences seen in Fig.~\ref{comparison_2} with respect to our new data are hard to assess. 


\section{\textit{R}-MATRIX ANALYSIS}
\label{Rmatrix} 
In this section we present an analysis of the differential cross section for $^4$He($^3$He,$^3$He)$^4$He elastic scattering using $R$-matrix theory~\cite{Lane_1958}. The phenomenological $R$-matrix code {\sc{Azure2}}~\cite{AZURE2} is used to analyze the elastic scattering data from this experiment and from Barnard \textit{et al}.~\cite{Barnard_1964}. We adopt the alternative parametrization of $R$-matrix theory presented in Ref.~\cite{brune2002}, so the $R$-matrix parameters are expressed in terms of the observed resonance energy $\tilde{E}$ and the observed reduced width amplitude $\tilde{\gamma}$. The channel radius is fixed at 4.2 fm. A channel radius of 4.3 fm was adopted in Ref.~\cite{deBoer14}.
\par
Most of the experiments reported in the literature are studies of the structure
of $^7$Be and consequently are focused on higher energies than the present work~\cite{Phillips_1958, Tombrello_1963, Barnard_1964, Spiger_1967, Ivanovich_1967, Chuang_1971, Hardy_1972, Boykin_1972}. The data of Barnard \textit{et al}.~\cite{Barnard_1964} were found to contain the most complete uncertainty information. The data of Spiger and Tombrello~\cite{Spiger_1967} are reported to have a systematic uncertainty as low as 1.1$\%$ and a maximum relative error of 9$\%$. The Spiger and Tombrello measurement only extends as low as $E$ [$^3$He]=4.7 MeV. The only measurement extending to lower energies, that of P. Mohr \textit{et al}.~\cite{Mohr_1993}, does not quantify systematic uncertainties and, as shown in Fig.~\ref{comparison_2}, has unexplained systematic variations of the cross section with angle. For these reasons, only the data of Barnard \textit{et al}.~\cite{Barnard_1964} and the data reported in this work were used in the $R$-matrix analysis. All data included in the analysis were taken at energies below the proton emission threshold.  
\par
\begin{figure}[btp]
    \centering
    \includegraphics[width=1.0\columnwidth,angle=0]{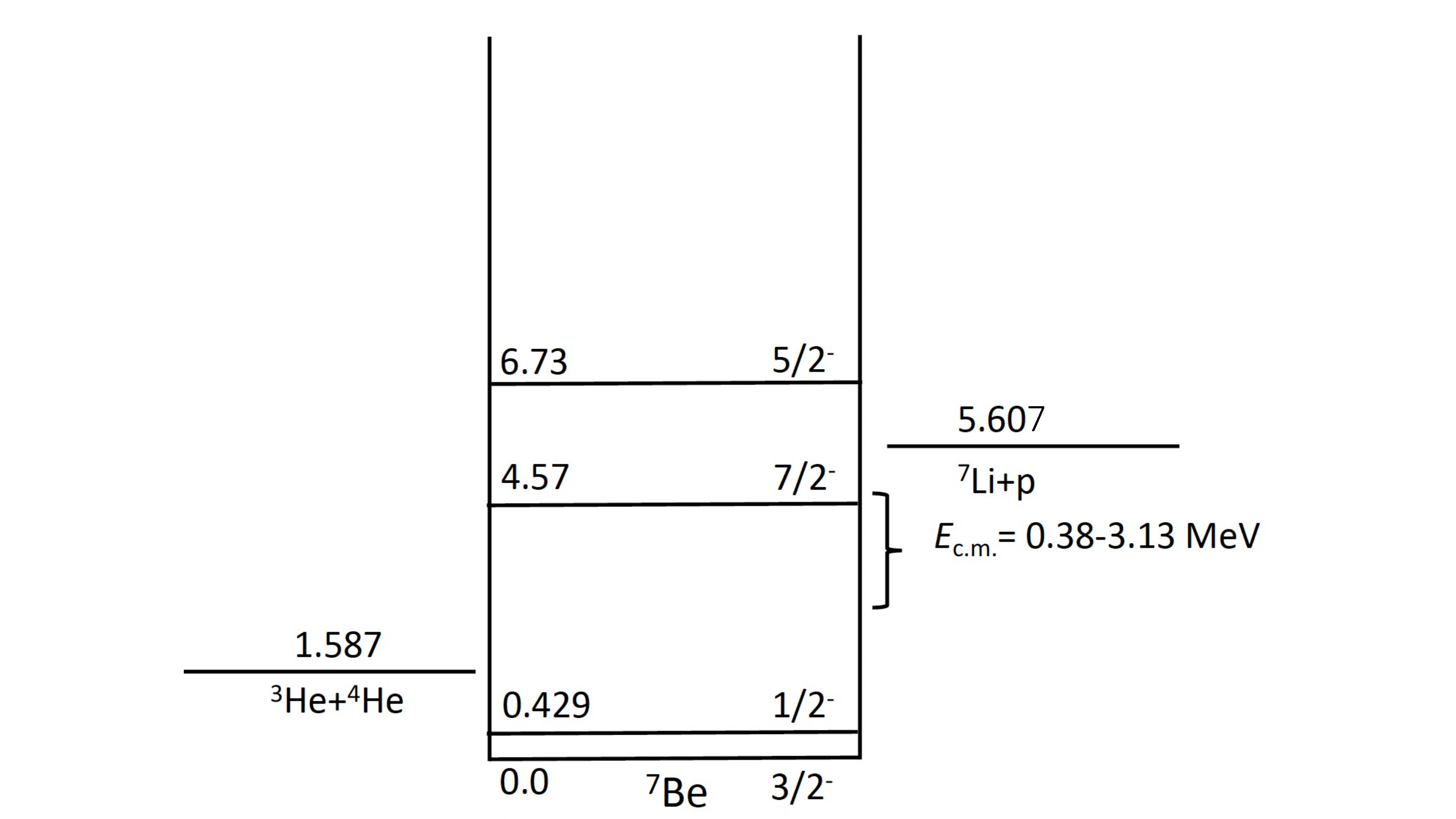}
    \caption{The levels and separation energies introduced in the $R$-matrix fit were taken from Ref.~\cite{NNDC}.  The energy range covered in the present measurement is represented by the curly brace. All energies are in MeV.}
    \label{level_diagram}
\end{figure}

The ground state spins and parities of $^4$He and $^3$He are 0$^+$ and 1/2$^+$, respectively. Restricting our calculations to orbital angular momentum $\ell\leq3$, the allowed total angular momentum and parities in $^7$Be are 1/2$^+$, 1/2$^-$, 3/2$^-$, 5/2$^+$, 3/2$^+$, 7/2$^-$, and 5/2$^-$.  The level diagram of the compound nucleus $^7$Be is shown in  Fig.~\ref{level_diagram}, with the energies of the levels taken from Ref.~\cite{NNDC}.  The energy range covered in this experiment, $0.38 \le E_{\text{c.m.}} \le 3.13$~MeV, is highlighted. The $R$-matrix analysis was started with the states of $^7$Be shown in Fig.~\ref{level_diagram}.  But, within the experimental energy range, the states 1/2$^+$, 5/2$^+$, and 3/2$^+$ are not identified in the literature. Therefore, these channels are introduced into the analysis via background levels. We also add background levels in the 1/2$^-$ and 3/2$^-$ channels, in addition to the levels that represent the bound states of ${}^7$Be which exist in those channels. The background levels for the 5/2$^+$, and 3/2$^+$ states are introduced at an excitation energy of 12.0 MeV. The excitation energies of the 1/2$^+$, 1/2$^-$, and 3/2$^-$ background levels are fixed at 14.0, 21.6, and 21.6 MeV, in order to reproduce the trend of the Spiger and Tombrello~\cite{Spiger_1967} phase shifts for $s$- and $p$-waves at high energies.

Similarly, the alpha width of the 5/2$^-$ state is fixed at 1.9~MeV to reproduce the trend of the Spiger and Tombrello~\cite{Spiger_1967} 5/2$^-$ phase shift data up to $E$[$^3$He]=9 MeV. 
The asymptotic normalization constants (ANCs) used for the sub-threshold 3/2$^-$ and 1/2$^-$ states are fixed at $C_{3/2}$=3.7 fm$^{-1/2}$ and $C_{1/2}$=3.6 fm$^{-1/2}$~\cite{deBoer14}, respectively. Including radiative capture data provides better constraints on the ANCs than can be obtained from scattering data alone, so we chose to fix the ANCs.

\subsection{Bayesian Analysis}
\label{bayesian_analysis}

In what follows we did not use AZURE2's built-in $\chi^2$ analysis, instead employing 
 a Markov Chain Monte Carlo (MCMC) analysis to obtain probability distributions and study how parameter uncertainties propagate to extrapolated quantities.
The goal of this MCMC analysis is to approximate the Bayesian posterior distribution function for the $R$-matrix parameters, which we collect into a vector $\vec{\theta}$~\footnote{We use $\vec{\theta}$ to represent the multi-dimensional parameter vector, trusting the reader to distinguish this from a scattering angle using context.}. By Bayes's theorem that posterior can be obtained as
\begin{equation}
    \label{eq:bayes_theorem}
    p(\vec{\theta}|D) = \frac{p(D|\vec{\theta})p(\vec{\theta})}{p(D)}~,
\end{equation}
where the likelihood, $p(D|\vec{\theta})$, is chosen to be
\begin{equation}
    \label{eq:likelihood}
    p(D|\vec{\theta}) = \prod_{i,j}\frac{1}{\sqrt{2\pi}\sigma_{i,j}}  e^{-\frac{1}{2}\chi^2_{i,j}}~,
\end{equation}
where 
\begin{equation}
\chi^2_{i,j}=\frac{(f(x_{i,j})-\tilde{c}_{i,j} y_{i,j})^{2}}{(\tilde{c}_{i,j} \sigma_{i,j})^2},
\label{eq:chi-square0}
\end{equation} 
where~\footnote{The following description applies only to the SONIK data, the Barnard data is treated via one overall normalization factor. For Barnard data, the overall systematic normalization, $c_{\text{Barnard}}$, the prior we set is a Gaussian with mean 1 and standard deviation 0.05, in accord with Ref.~\cite{Barnard_1964}.} $i$ indexes the beam energy, $j$ indexes the detector,  $f(x_{i,j})$ is the differential scattering cross section from the $R$-matrix, $y_{i,j}$ is the data point value, $\sigma_{i,j}$ is the statistical uncertainty of the data point, and $\tilde{c}_{i,j}$ is a composite normalization factor defined by Eq.~(\ref{eq:cij}) below.

The prior distribution function
\begin{equation}
    \label{eq:prior}
    p(\vec{\theta}) = \prod_{i=1}^{N_p} p(\theta_i)~,
\end{equation}
is a product of each parameter's prior distribution, with $N_p$ representing the number of sampled parameters.
Typically, with $\chi^2$ minimization, each parameter is allowed to move freely in an unbounded space.
The analyses presented here mimic that freedom by imposing uniform priors with generous upper and lower bounds.
Those bounds are shown in Table~\ref{tab:prior_limits}; these priors are quite similar to the ones adopted in Ref.~\cite{Odell:2021nmp}. Since the analysis we perform here only considers elastic scattering data, the sign of the reduced width amplitudes cannot be uniquely determined. We obtain a solution where all partial widths have the same signs found in Ref.~\cite{Odell:2021nmp}, where capture data was also included.

\begin{table}[]
    \centering
    \begin{tabular}{c|c|c}
         Parameter & Lower Bound & Upper Bound \\
         \hline
         $\Gamma_{\alpha}^{(1/2-)}$ & -150 {\rm MeV} & 150 {\rm MeV} \\
         $\Gamma_{\alpha}^{(1/2+)}$ & 0 MeV & 100 MeV \\
         $\Gamma_{\alpha}^{(3/2-)}$ & -100 MeV & 100 MeV \\
         $\Gamma_{\alpha}^{(3/2+)}$ & 0 MeV & 100 MeV \\
         $\Gamma_{\alpha}^{(5/2+)}$ & 0 MeV & 100 MeV \\
         $E_x^{(7/2-)}$ & 2 MeV & 10 MeV \\
         $\Gamma_{\alpha}^{(7/2-)}$ & 0 & 10 MeV
    \end{tabular}
    \caption{Limits of uniform prior distributions set for the $R$-matrix parameters that were sampled in the MCMC analysis.}
    \label{tab:prior_limits}
\end{table}

The normalizations applied to the data, $c_{i,j}$, are, for each data point, a product of three different effects: $c_{\rm SONIK}$, $c_i$, and $c_j$. The overall systematic uncertainty is accounted for with $c_{\rm SONIK}$. This factor is applied to all data points. An energy-dependent systematic uncertainty is accounted for with $c_i$. Each energy bin has its own associated $c_i$, which applies to the data from all three interaction regions at a given energy. A detector-specific systematic uncertainty is accounted for with $c_j$. All points measured with the same detector are adjusted by $c_j$. There are 27 detectors in total. The resulting normalization adjustment of the data is then 
\begin{equation}
    \tilde{c}_{i,j}=c_{\rm SONIK} c_i c_{j},
    \label{eq:cij}
\end{equation}
where $c_i$ and $c_j$ depend on the energy bin and detector used to measure $y_{i,j}$. We set Gaussian priors for each of three types of normalizations. For the overall systematic normalization, $c_{\rm SONIK}$, the prior we set is a Gaussian of mean 1 and standard deviation 0.02, in accord with the error budget of Table~\ref{systematic_uncertainty_1}. For the energy bin normalizations, $c_i$, the prior chosen is a Gaussian of mean 1 and standard deviation $\sigma_{E_i}$, with the $\sigma_{E_i}$'s are tabulated in the fourth column in table~\ref{chi_square}. Finally, the detector-specific normalizations, $c_j$, are each assigned a Gaussian prior of mean 0.96 and standard deviation 0.032, to account for variances in the aperture dimensions, cf. Sec.~\ref{G-factor}.

Since the priors on the $R$-matrix parameters are intentionally left very broad, our Bayesian posterior can be well approximated as a likelihood $e^{-\chi^2_{\rm aug}/2}$, where the chi-squared is:
\begin{equation}
    \label{eq:chi-square}
    \chi^2 = \sum_{i}\left(\sum_{j} \frac{(f(x_{i,j})-\tilde{c}_{i,j} y_{i,j})^{2}}{(\tilde{c}_{i,j} \sigma_{i,j})^2}\right)
\end{equation}
and the augmented chi-squared is:
\begin{widetext}
\begin{equation}
        \label{eq:chi-square-aug}
        \chi_{\rm aug}^2 = \chi^2 + \sum_i \left(\frac{c_{i}-1}{\sigma_{E_i}}\right)^2 +
        \sum_j \left(\frac{c_j-0.96}{0.032}\right)^2 + \left(\frac{c_{\rm SONIK}-1}{0.02}\right)^2 + \left( \frac{ c_{\rm Barnard} - 1 } { 0.05 } \right)^2~.
    \end{equation}
\end{widetext}

The MCMC analysis was performed with a publicly available ensemble sampler, emcee~\cite{emcee}, and the previously mentioned $R$-matrix code, AZURE2.
The pairing was enabled by a publicly available Python layer, BRICK~\cite{Odell:2021nmp}.




\section{ANALYSIS USING HALO EFFECTIVE FIELD THEORY}
\label{eft}
An EFT is a controlled expansion in a ratio $Q \equiv \frac{p_{\text{typ}}}{\Lambda}$, where $p_{\text{typ}}$ is the low-momentum scale that typifies the scattering 
and $\Lambda$ is the momentum scale at which the theory breaks down---see, e.g., Ref.~\cite{kaplan1995effective} for an introduction.  
The EFT expansion of an observable $y$ in powers of $Q$ can be written as \cite{weinberg1979phenomenological, epelbaum2009modern, wesolowski2019exploring}
\begin{equation}
    \label{power_counting_eqn}
    y(p,\theta) = y_{\text{ref}}(p,\theta) \sum_\nu c_\nu(p,\theta) \ Q^\nu.
\end{equation}
Here $\nu$ indexes the order of different contributions. We denote $\nu=0$ as leading order (LO), $\nu=1$ next-to-leading order (NLO), and $\nu=2$ as next-to-next-to-leading order (NNLO). 

In this section we briefly describe the EFT  we use to calculate the ${}^3$He-$\alpha$ scattering reaction. Full details of the EFT can be found in Ref.~\cite{Poudel_2022}.

 This EFT is built on the scale separation between the large de Broglie wavelength of the quantum-mechanical scattering process and the small size of the ${}^3$He and $\alpha$ nuclei. It is thus an example of ``Halo EFT"; see Ref.~\cite{Hammer:2017,Hammer:2019} for recent review. In this approach $^7$Be is a bound state of $^3$He and $\alpha$ nuclei. Such a description is accurate because the energies by which the ground- and excited-state of $^7$Be are below the ${}^3$He-$\alpha$ scattering threshold are 1.6 and 1.2 MeV \cite{tilley2002energy} respectively. These are small compared to the energy scales at which ${}^3$He and ${}^4$He can be broken up into smaller constituents. These energy scales, as well as the sizes of the two helium isotopes, yield an EFT breakdown momentum of $\Lambda \approx 200$ MeV/c~\cite{zhang2020s}. 
 
 We take the typical momentum of the collision to be $p_{\rm typ}=\max\{q,p\}$, where  $q=2p\sin(\theta/2)$ is the momentum transfer of the scattering reaction. The bulk of the SONIK data were taken for $p$ between 60 MeV/c and 90 MeV/c.  In this energy range Halo EFT has been successfully applied to the $^3$He($\alpha$,$\gamma$)$^7$Be reaction~\cite{higa2018radiative,zhang2020s,Rupak_2020} and used to fit scattering phase shifts~\cite{higa2018radiative,Rupak_2020}. In Ref.~\cite{Poudel_2022} the choice $\Lambda=200$ MeV/c was validated by showing that it leads to an EFT with a regular convergence pattern, i.e., once $Q$ is chosen in this way the coefficients $c_0$, $c_1$, and $c_2$ in Eq.~(\ref{power_counting_eqn}) have roughly the same size. 
 
The EFT contains only minimal assumptions about the ${}^3$He-$\alpha$ dynamics: rotational invariance, unitarity, analyticity of the amplitude, and the presence of a short-range strong interaction as well as a long-range Coulomb potential. Since the Coulomb-modified effective-range expansion (CM-ERE) is based on the same set of assumptions the Halo EFT $t$-matrix has the same form as that obtained in the CM-ERE ~\cite{Kong:1999,Higa:2008,Poudel_2022}. The EFT Lagrangian is expressed as an expansion in powers of $p^2$, so, at a given order in the EFT, the CM-ERE is reproduced up to the corresponding order of $p^2$. 

\begin{widetext}

The CM-ERE amplitude  associated with ${}^3$He-$\alpha$ scattering in the $(l,J=l\pm \frac{1}{2})^\text{th}$ channel, $T_l^{\pm}$, takes the form~\cite{bethe1949theory,hamilton1973coulomb}:
\begin{equation}
\label{t_matrix_PWA}
    T^\pm_l(E + i \epsilon)= -(2l+1)\frac{2\pi}{\mu}
    \frac{\Bigg{[} \frac{\Gamma(2l+2)}{2^l \Gamma(l+1)} \Bigg{]}^2 C_l^2(\eta) \text{e}^{2i\sigma_l} p^{2l}P_l(\cos \theta)}{\Bigg{[} \frac{\Gamma(2l+2)}{2^l \Gamma(l+1)} \Bigg{]}^2 C_l^2(\eta) p^{2l+1} (\cot \delta^\pm_l-i)},
\end{equation}
where the quantities $\delta^\pm_l$ are the phase shifts for the channels $l^\pm$. 
The phase shift for the $\pm$ channels in the $l$th partial wave are, in turn, given by 
\begin{equation}
    \label{CMERE}
    \Bigg{[}\frac{\Gamma(2l+2)}{2^l \Gamma(l+1)} \Bigg{]}^2 C^2_l(\eta) p^{2l+1} (\cot \delta^\pm_l-i) = 2k^{2l+1}_c K^\pm_l(E)
    -\frac{2k_c p^{2l}}{(\Gamma(l+1))^2} \frac{\Gamma(1+l+i\eta)\Gamma(1+l-i\eta)}{\Gamma(1+i\eta)\Gamma(1-i\eta)} H(\eta),
\end{equation}
\end{widetext}
where
\begin{equation}
    \label{K-function}
    K^\pm_l = \frac{1}{2k_c^{2l+1}}\Bigg{(}-\frac{1}{a^\pm_l} + \frac{1}{2} r^\pm_l p^2 + \frac{1}{4} P^\pm_l p^4 + O((p^2)^3) \Bigg{)}
\end{equation}
is the effective-range function.
Equations \eqref{CMERE} and \eqref{K-function} relate the phase shifts to the coefficients of powers of $p^2$ in the expansion of the function $K$. $K(E)$ is analytic in $p^2$ for $|p|< 1/R$, where $R$ is the range of the strong interaction.  The coefficients of a Taylor series expansion of $K$ in $p^2$ are (apart from numerical factors) the effective range parameters (ERPs). To obtain phase shifts from ERPs, the polynomial $K$-function is truncated at a suitable order. In equations~\eqref{t_matrix_PWA} and~\eqref{CMERE}, $P_l(\cos \theta)$ is the $l$th Legendre polynomial calculated at the cosine of scattering angle $\theta$ and the quantities $C_l(\eta)$ and $e^{2i\sigma_l}$ are given respectively by 
\begin{equation}
    \label{coulomb_phase_shift}
    \text{e}^{2 {\rm i}\sigma_l}= \frac{\Gamma(1+l+{\rm i}\eta)}{\Gamma(1+l-{\rm i}\eta)}
\end{equation}
and
\begin{equation}
    \label{CL}
    C_l(\eta) = \frac{2^l}{\Gamma(2l+2)}\exp{(-\pi \eta/2)}\ |\Gamma(1+l+{\rm i}\eta)|.
\end{equation}

\begin{table}[b]
    \centering
    \caption{Hierarchy of power counting in our EFT:}
    \begin{tabular}{c c c c}
     & s-wave & p-wave & $\nu$ \\
     \hline \hline \\
    LO & - & - & 0\\
    NLO & $r_0$ & $r^+_1$, $P^\pm_1$ & 1\\
    NNLO & $\frac{1}{a_0}$ & $\frac{1}{a^\pm_1}, r^-_1$ & 2  
    \end{tabular}
    \label{power_counting_table}
\end{table}
The EFT power counting is a particular assignment of the terms that should appear in the $K$-function at a given order $\nu$. That assignment is chosen to ensure that the pattern (\ref{power_counting_eqn}) is satisfied. In Ref.~\cite{Poudel_2022} an assignment that achieves this was found although it should be clear that such an assignment depends on having some knowledge of the size of the ERPs themselves, and so can only be accomplished in the light of at least some data on the system.

The organization of Ref.~\cite{Poudel_2022} has at LO only the contributions proportional to the $H$-function for both $s$-and $p$-wave channels, i.e., it takes $K=0$. We say this piece of the inverse amplitude is of order $p$ for $s$-waves and of order $p^3$ for $p$-waves. (We assume $\eta \sim 1$.)
The terms proportional to ERPs that appear in $K$ are corrections to this limit. 

For $s$-waves, the term $\frac{1}{a_0}$ is a very small momentum and we take it to be $\sim p_{\text typ}^3/\Lambda^2$. Meanwhile, the $s$-wave effective range $r_0$ scales naturally $\sim \frac{1}{\Lambda}$. Therefore for $s$-waves, we include the term proportional to $r_0$ in the effective-range expansion  at NLO (its effect is $\sim p_{\text typ}^2/\Lambda$) and that from $\frac{1}{a_0}$ at NNLO. 
Regarding $p$-waves, we consider both $p$-wave shape-parameter terms, $\frac{1}{4} P^\pm_1 p^4$, at NLO, as the shape parameters are natural ($\sim 1/\Lambda$) and so this term is $\sim p_{\text{typ}}^4/\Lambda$. The other $p$-wave ERPs are unnaturally small, so in the $J=\frac{3}{2}^-$ channel, we take the contribution from $\frac{1}{2}r_1^+ p^2$ at NLO and consider that of $\frac{1}{a_1^+}$ only at NNLO. Both $\frac{1}{a_1^-}$ and $\frac{1}{2}r_1^- p^2$ are considered to be NNLO effects. This organization is summarized in Table~\ref{power_counting_table}. All results presented below are computed at NNLO.

However, even at NNLO, if we are to describe the higher-energy portion of the SONIK data at the required accuracy, we must include the  ${\frac{7}{2}}^-$ partial wave in the analysis. To account for the impact of the  ${\frac{7}{2}}^-$ ${}^7$Be level at $E_{\rm x}=4.57$ MeV \cite{Piluso_1971}  on observables in the energy range of interest Ref.~\cite{Poudel_2022} employed a phenomenological treatment of it, based on {\it R}-matrix theory~\cite{Lane_1958}. The focus both in Ref.~\cite{Poudel_2022} and here is not on the resonance itself. The goal of phenomenologically adding its amplitude to the EFT analysis is solely to stop it contaminating the extraction of the ERPs. Since Ref.\cite{Poudel_2022} points out that the inclusion of contributions to the amplitude from the $\frac{5}{2}^-$ partial wave is also essential for a consistent analysis of the data, we also include them in the EFT analysis presented here. In adding these contributions to the scattering amplitude from $f$-waves, we employ the following resonance energies and physical widths:
\begin{eqnarray*}
E^{\frac{7}{2}^-}_R &=& 5.22 \ {\rm MeV \ in \ lab}, \\
E^{\frac{5}{2}^-}_R &=& 9.02 \ {\rm MeV \ in \ lab}, \\
\Gamma^{\frac{7}{2}^-} &=& 0.159 \ {\rm MeV}, \\
{\rm and} \ \Gamma^{\frac{5}{2}^-} &=& 1.8 \ {\rm MeV}.
\end{eqnarray*}
which are used to generate the $\frac{7}{2}^-$ and $\frac{5}{2}^-$ phase shifts using $l=3$ in the following formula:
\begin{equation}
    \label{eq:r_matrix_phase_shift}
    \delta_{l^\pm} = -\phi_l + \tan^{-1}\frac{P_l(E, \rho)}{P_l(E^c_R, \rho)}\frac{\frac{1}{2}\Gamma^{c}}{E^c_R-E}.
\end{equation}
In Eq.~(\ref{eq:r_matrix_phase_shift}),
\begin{equation}
    \label{eq:hard_sphere_phase_shift}
    \phi_l=\tan^{-1}\frac{F_l(\eta, \rho)}{G_l(\eta, \rho)}
\end{equation}
and
\begin{equation}
    \label{eq:penetration}
    P_l(E, \rho) = \rho/(F^2_l+ G^2_l)
\end{equation}
where $F_l$ and $G_l$ are the usual regular and irregular Coulomb functions~\cite{abramowitz1964handbook} and $\rho = pr$ where $r$ is the channel radius, here taken to be $4.2$ fm. The channels are represented with superscript $c$; for $l=3$, $c=\frac{7}{2}^-, \frac{5}{2}^-$.  
The resonance energies and physical widths that we adopt mimic the f-wave phase shifts produced by Bayesian {\it R}-matrix analysis.  

The ERPs---$a_0, r_0, a^+_{1}, r^+_{1}, P^+_{1}, a^-_{1}, r^-_{1} \ \text{and} \ P^-_{1}$---span an 8-dimensional parameter space. Using relationships between the effective-range amplitude and bound-state properties~\cite{zhang_2020,Poudel_2022}, we
reparametrize the space in terms of the ANCs, replacing $r^\pm_{1}$ by $C^\pm_1$ using equation (43) from Ref.~\cite{Poudel_2022}. The ANCs are fixed at $C^+_1=C_{3/2}=3.7~{\rm fm}^{-1/2}$ and $C^-_1=C_{1/2}=3.6~{\rm fm}^{-1/2}$---as was done in the $R$-matrix analysis. We also determine $a^\pm_{1}$ from the location of the two ${}^7$Be bound states using equation (39) from Ref.~\cite{Poudel_2022}. These two constraints reduce the eight-dimensional ERP space to a four-dimensional one.

To compute the posterior of the EFT parameters we employ Bayes's theorem, Eq.~(\ref{eq:bayes_theorem}), as was done in Sec.~\ref{bayesian_analysis} for the $R$-matrix analysis. However, for the EFT analysis we employ a $\chi^2$ function that is different to the standard one,  Eq.~\ref{eq:chi-square}.  Truncation of the EFT series at order $\nu_{max}$ induces an error in the observable $y$~\cite{Furnstahl:2016} at data point $j$ in data set $i$ of:
\begin{equation}
    \Delta y(x_{i, j})=y_{\rm ref}(x_{i,j}) c^{\rm rms} Q_{i, j}^{\nu_{\rm max} + 1},
    \label{eq:errormodel}
\end{equation}
where $c^{\rm rms}$ represents the rms value of the EFT coefficients defined in Eq.~(\ref{power_counting_eqn}) and $x_{i, j}$ is a kinematic point $(p_j, \theta_j)$ in dataset $i$. We therefore
use a modified $\chi^2$~\cite{wesolowski2019exploring}:
\begin{equation}
    \label{eq:chi-square2}
        \chi^2_{{\rm EFT},i} = [\vec{r}^{\, T}(\sigma^{\rm expt} + \sigma^{\rm th})^{-1}\vec{r}]_i 
\end{equation}
where $\chi^2_{{\rm EFT},i}$ is the modified $\chi^2$ of dataset $i$. The matrix elements of the theory covariance matrix in  dataset $i$ are 
\begin{equation}
    \label{theory_cov}
    \sigma^\text{th}_{jk}=(y_{\text{ref}})_j (y_{\text{ref}})_k (c^{\rm rms})^2 Q^{\nu_{\rm max}+1}_j Q^{\nu_{\rm max}+1}_k.
\end{equation}
It accounts for the error due to omitted higher-order terms in the EFT. In this analysis we take that error to be completely correlated across the kinematic space, see Ref.~\cite{Melendez:2019} for a more advanced treatment. We take the experimental covariance matrix to be diagonal
\begin{equation}
  \sigma^{\rm expt}_{jk}=\tilde{c}_{i,j}^2 \sigma^2_j\delta_{jk},
\end{equation}
while the entries of the residual vector $\vec{r}$ for dataset $i$ are defined by
\begin{equation}
r_j=f(x_{i, j})-\tilde{c}_{i,j} y_{i, j}.    
\end{equation}
The inclusion of the truncation errors modifies the likelihood to
\begin{equation}
    \label{eq:likelihood2}
    p(D|\vec{\theta}) = \frac{1}{\sqrt{(2\pi)^N {\rm det}(\sigma^{\rm expt}+\sigma^{\rm th})}}  e^{-\frac{1}{2}\sum_{i}\chi^2_{{\rm EFT},i}}.
\end{equation}
In Eq.~\ref{eq:likelihood2} $N$ is the total number of data points and det means the matrix determinant.

 In addition to the ERPs, Eq.~(\ref{eq:chi-square2}) includes as parameters the normalizations $\tilde{c}_{i,j}$ of the differential cross-section data, each of which is a product of the three different normalization factors, see Sec.~\ref{bayesian_analysis} for details. We adopt the priors for the overall, energy-dependent, and detector-dependent normalization factors specified in Sec.~\ref{bayesian_analysis}.

To construct the EFT error model we take $\nu_{max}=2$, since the  calculation is carried out to NNLO. Meanwhile, $(y_{\text {ref}})_j$ is taken to be the LO cross section at data point $j$.  $\bar{c}$ is then estimated from the size of the shifts from LO to NLO and NLO to NNLO to be $c^{\rm rms}=0.70$, as described in Ref.~\cite{Poudel_2022}. Further details regarding the Bayesian analysis of the NNLO Halo EFT calculation can also be found in that work. 

We observe that the EFT has an expansion parameter of 0.2 at forward angles in the lowest SONIK energy bin, but $Q$ approaches one for the backward-angle data at the highest SONIK $E_{\rm c.m.}$ of 3.1 MeV. We therefore do not expect Halo EFT to accurately describe all the data collected in this experiment. That, after all, is why Poudel and Phillips truncated their analysis at $E_{\rm c.m.}=2.5$ MeV~\cite{Poudel_2022}. The inclusion of the truncation error in the likelihood ameliorates the theory's failure to describe higher-$Q$ data, since it decreases the statistical weight of data for which $Q$ is larger. Nevertheless,
the sensitivity of the inference to assumptions regarding the nature of the truncation error becomes quite severe as $Q \rightarrow 1$. In spite of this, we will include all the SONIK data in our Halo EFT analysis, so that we can make a direct comparison with the $R$-matrix analysis.



\section{RESULTS}
\label{results}
\subsection{{\em R}-matrix results}

\begin{figure}
\centering
\includegraphics[width=\columnwidth]{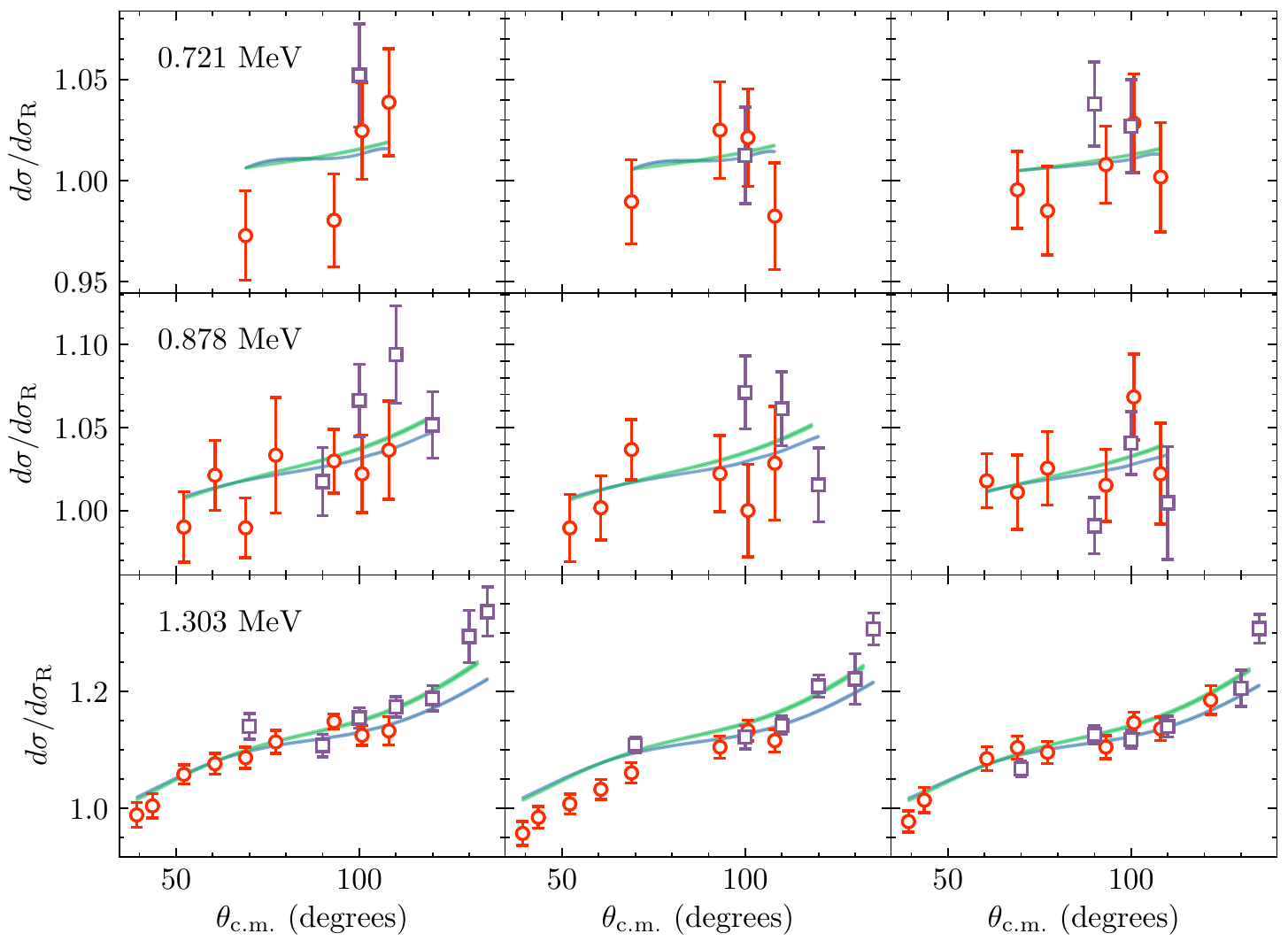}
\caption{Differential elastic scattering cross sections, relative to Rutherford's prediction, as measured with SONIK. Results are obtained from an MCMC analysis of SONIK and Barnard data. Bands encompass the 16th to 84th percentile of the inferred probabibilty distributions. Blue bands correspond to the $R$-matrix analysis, and green bands correspond to the Halo EFT analysis. Red circles with error bars indicate ${}^3{\rm He}$ peaks. Purple squares with error bars indicate ${}^4{\rm He}$ peaks. The three panels along a row for a given $E [^{3}$He] beam energy are from interaction regions I, II, and III, respectively.}
\label{Fits_SONIK_1}
\end{figure}

\begin{figure}
    \centering
    \includegraphics[width=\columnwidth]{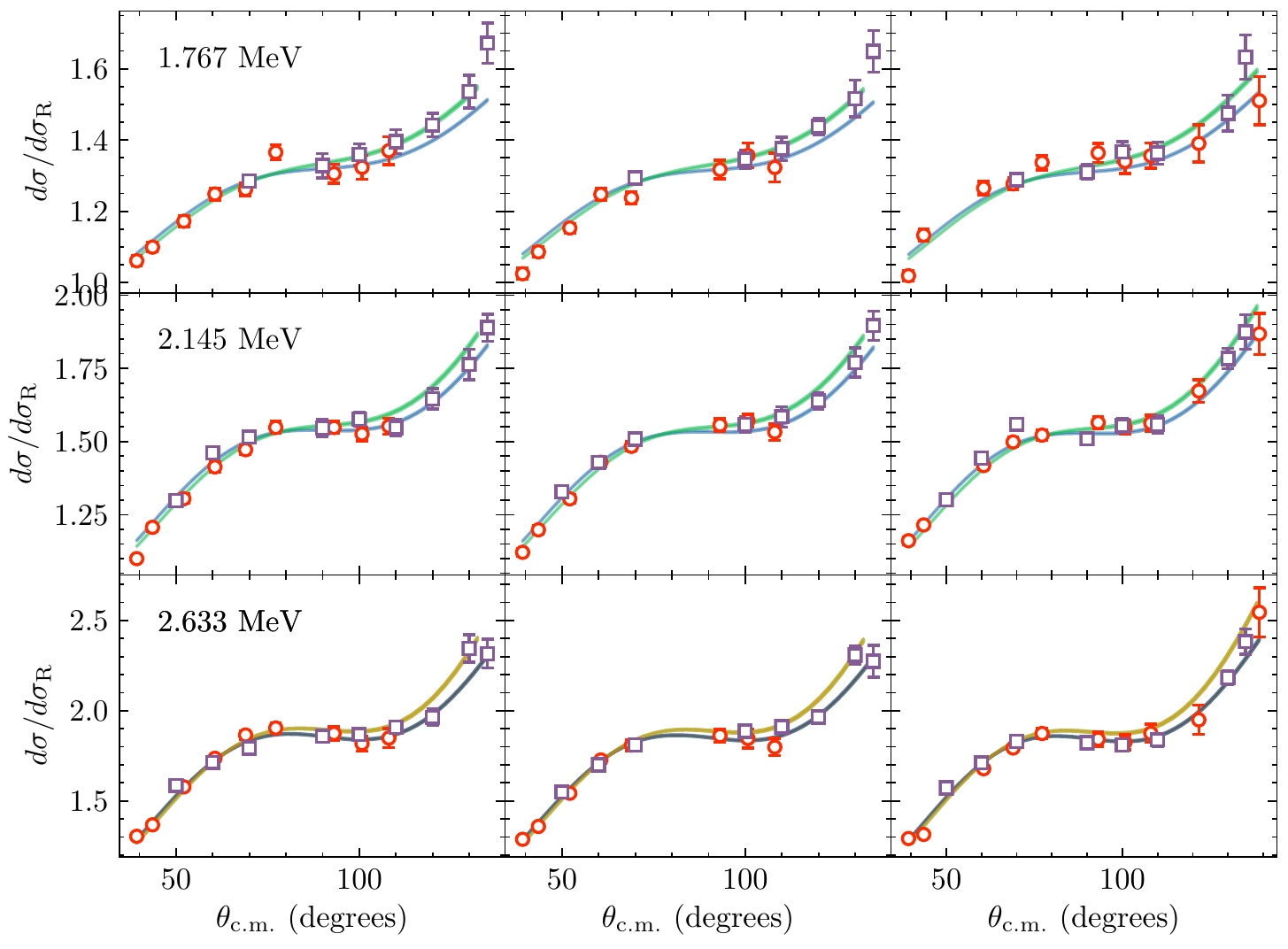}
    \caption{Differential elastic scattering cross sections, relative to Rutherford's prediction, as measured with SONIK. Colors and symbols are as described in Fig.~\ref{Fits_SONIK_1}.
    The additional yellow band in the bottom row corresponds to the second run at $E[{}^3{\rm He}]=2.633 {\rm MeV}$ discussed in \ref{sec:beam_normalization}.}
    \label{Fits_SONIK_2}
\end{figure}

\begin{figure}
    \centering
    \includegraphics[width=\columnwidth]{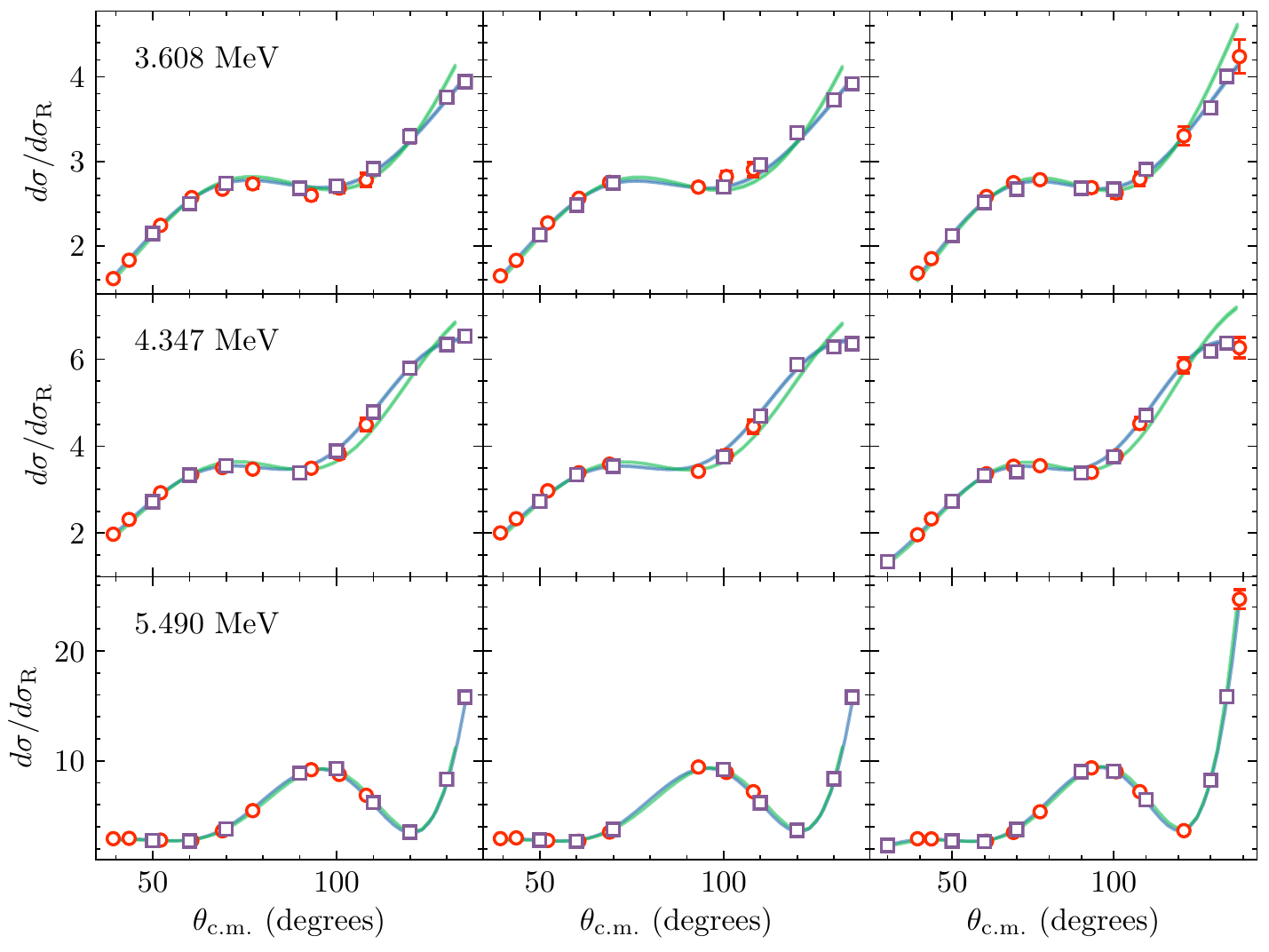}
    \caption{Differential elastic scattering cross sections, relative to Rutherford's prediction, as measured with SONIK. Colors and symbols are as described in Fig.~\ref{Fits_SONIK_1}.}
    \label{Fits_SONIK_3}
\end{figure}

\begin{figure}
    \centering
    \includegraphics[width=\columnwidth]{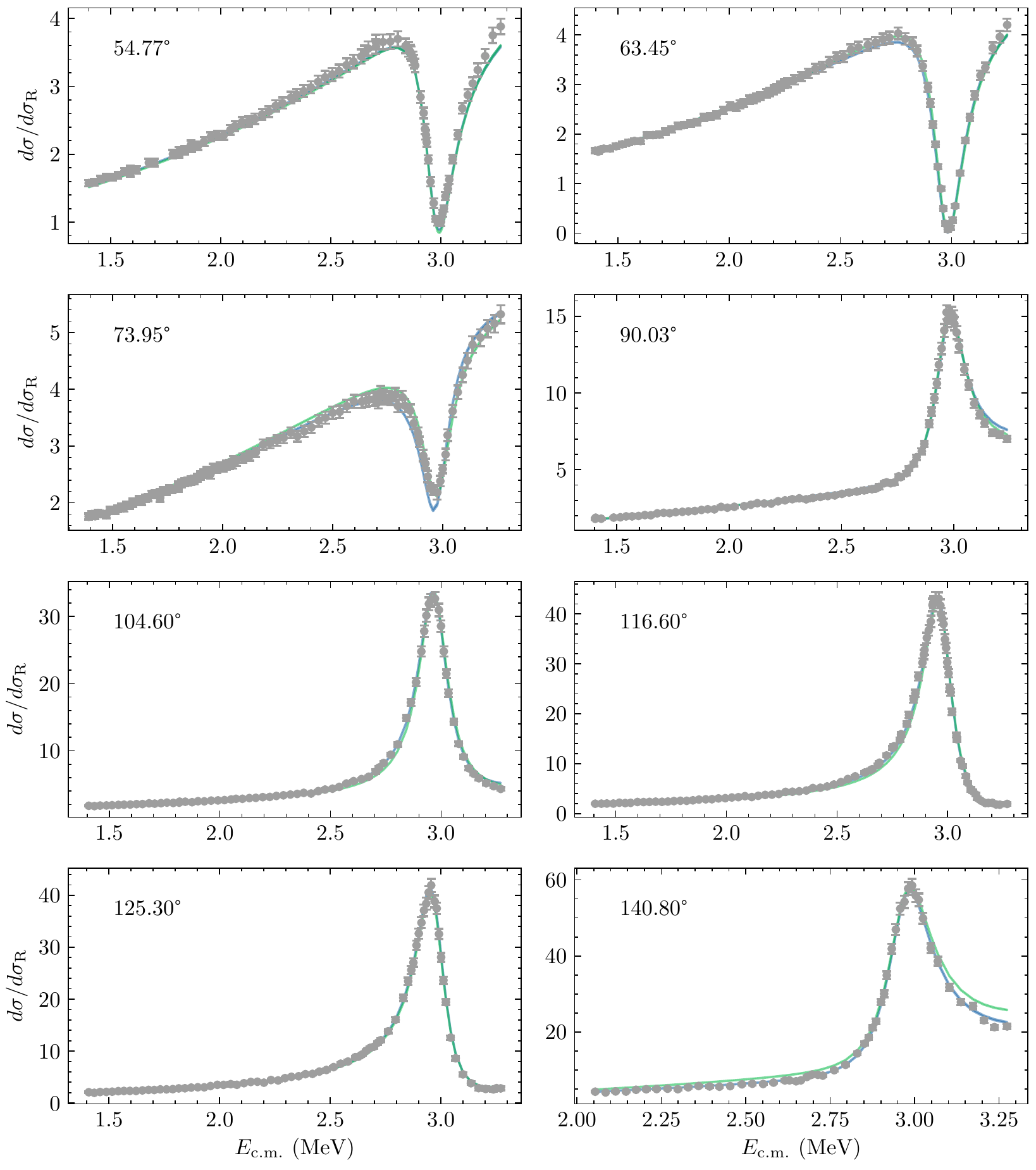}
    \caption{Differential elastic scattering cross sections, relative to Rutherford's prediction, as reported in Ref.~\cite{Barnard_1964}. Results are obtained from an MCMC analysis of SONIK and Barnard data. Bands are as in Fig.~\ref{Fits_SONIK_1} and grey circles represent the data from Ref.~\cite{Barnard_1964}. Total $\chi^2$ at maximum posterior probability for the $R$-matrix fit is 1098.56, resulting in $\chi^2/{\rm datum}$ = 1.70. The total $\chi^2$ for the EFT fit to these data is 1996.24.}
    \label{Fits_barnard}
\end{figure}

The results from the simultaneous fitting of the elastic scattering data of the current measurement and the data of Barnard \textit{et al}.~\cite{Barnard_1964} are shown in Figs.~\ref{Fits_SONIK_1}, ~\ref{Fits_SONIK_2}, \ref{Fits_SONIK_3}, and \ref{Fits_barnard}. The blue bands in the figures correspond to
the $R$-matrix analysis, and green bands correspond to the Halo EFT analysis.
The medians of the normalization factors, $\tilde{c}_{i,j}$ for the SONIK data and $c_{\rm Barnard}$ for the Barnard data, have been applied to the data in the figures.
Although the total $\chi^2$ from both analyses are comparable at the lowest three energies, as seen in Fig.~\ref{Fits_SONIK_1} and with comparable $\chi^2$ and $\chi^2_{\rm EFT}$ values from Table~\ref{chi_square}, the two analyses differ in the angular distribution. The two analyses also differ significantly in terms of $\chi^2$ at the three highest SONIK energies, cf. Table~\ref{chi_square} and discussion in Sec.~\ref{res:haloeft}. The two analyses produce similar fits at the intermediate SONIK energies.
\par
The $\chi^2$ values for each data segment are presented in Table~\ref{chi_square} and the best fit $R$-matrix parameters are presented in Table~\ref{R-matrix_param}. A total of 1097 data points were fitted simultaneously with 46 free $R$-matrix parameters. The fits to the whole data set have a minimum reduced $\chi^2$ of 1.85. This value is calculated at the point in parameter space that maximizes the posterior and therefore at the best values of the normalization parameters found by the sampler. It cannot be straightforwardly interpreted as a measure of the quality of the $R$-matrix fit, since the uncertainties of the normalization factors are not accounted for in the covariance matrix used to compute this standard $\chi^2$.
  The dominant contribution to $\chi^2$ comes from the differential scattering cross section data points at forward angles (22.5$^\circ \leq \theta_\text{lab} \leq 35 ^\circ$) in the laboratory frame of reference. 
The width of the 5/2$^-$ level introduces a significant change in the total $\chi^2$ for the $R$-matrix fit to the differential scattering cross section results at $E [^{3}\text{He}]=5.490$~MeV. The width of the 7/2$^-$ level is reported to be
175~$\pm$~7 keV~\cite{Piluso_1971} which is consistent with the width obtained fitting the data from this work alone. However, the width of the 7/2$^-$ level obtained from the simultaneous fit of the data of this work and Barnard \textit{et al.}~\cite{Barnard_1964} is significantly lower than the value reported in
Ref.~\cite{Piluso_1971}. In fact, if the stated energy uncertainty of $\pm$40 keV for data above 4 MeV in Barnard \textit{et al.}~\cite{Barnard_1964} is accounted for in the fit the central value of both $\tilde{E}_{7/2-}$ and $\tilde{\gamma}$ change by more than the uncertainty quoted in Table~\ref{R-matrix_param}. The difference in the alpha widths of  the 7/2$^-$ resonance needs to be resolved by future experiments. 

At low energies, the $^4$He($^3$He,$\gamma$)$^7$Be reaction primarily occurs through $E1$ external $s$-wave capture contributions~\cite{ChristyandDuck}. However, $d$-wave capture and internal contributions must also be considered. The internal $s$-wave part comes from the $J=1/2^{+}$ background level, which interferes with the external contribution to produce the large
capture cross section at low energies~\cite{deBoer14}. Therefore, the $s$-wave scattering length, $a_0$, is of particular importance. 
It is related to the $R$-matrix parameters via~\cite{Paneru} 
  \begin{equation}\label{scatt_length}
  a_{0}=-a\left[\frac{M_{cc}}{x^{2}K_{1}^{2}(x)}-\frac{2I_{1}(x)}{x^{2}K_{1}(x)}\right],
 \end{equation}
 where $M_{cc}=\mathbf{\tilde{\gamma}}^{T}_{c}\mathbf{\tilde{A}}\mathbf{\tilde{\gamma}}_{c}$, ${\mathbf{\tilde{A}}}$ is the level matrix as defined in Ref.~\cite{brune2002}, $c$ is the channel index, $a$ is the channel radius, $I_{1}(x)$ and $ K_{1}(x)$ are modified Bessel functions with $x=\left(8 Z_{1}Z_{2}e^{2}\mu a /\hbar^{2} \right)^{1/2}$, 
  $Z_{1}e$ and $Z_{2}e$ are the nuclear charges, $\hbar$ is the reduced Planck's constant, and $\mu$ is the reduced mass. 
   
\begin{table*}
\renewcommand{\arraystretch}{1.5}
\caption{${}^3$He beam energies, corresponding center-of-mass energy range, angular range (in the c.m. frame), normalization uncertainty, denoted $\sigma_{E_{i}}$, at each energy, normalization factors $c_{i}$ obtained in the $R$-matrix fit, $\chi^2$ from both $R$-matrix and Halo EFT, and number of data points $N$ of the angular distributions from the SONIK experiment reported in this work (part A) and the excitation function of Barnard \textit{et al}.~\cite{Barnard_1964} (part B). The $\chi^2$ and $\chi^2_{\rm EFT}$ per degree of freedom from $R$-matrix and Halo EFT analyses were found to be 1.85 and 3.15, respectively.} 
\begin{ruledtabular}
\begin{tabular}{cccccccc}
{$E~[^{3}\text{He}]$ (\text {MeV})}&{$E_{\text{c.m.}}$ (\text{MeV})} & \multicolumn{1}{c}{$\theta_{\text{c.m.}}$ (degrees)} &
\multicolumn{1}{p{2
cm}}{\centering Normalization  uncertainty} &
\multicolumn{1}{p{2 cm}}{\centering $c_i$ from \\ R-matrix \\ analysis} &
\multicolumn{1}{c}{$\chi^{2}_{{\rm R},i}$} & \multicolumn{1}{c}{$\chi^{2}_{{\rm EFT},i}$} &
\multicolumn{1}{c}{N}\\\hline
\multicolumn{2}{c}{\textbf{(A) This work}}& & &\\
$5.490$ & $3.122-3.127$&$30.00-138.90$ & 8.7$\%$ & $1.023^{+0.005}_{-0.005}$ & 171.00 & 214.23  & 53\\
$4.347$ & $2.470-2.476$ &$39.26-135.00$ &6.0$\%$ &  $0.975^{+0.004}_{-0.004}$ & 74.71 & 373.85 & 53\\
$3.608$ & $2.045-2.052$ &$39.26-135.00$ & 7.5$\%$ & $0.992^{+0.004}_{0.004}$ & 49.28 & 253.89  & 52\\
$2.633$ & $1.488-1.496$ &$30.00-138.90$ & 3.7$\%$ & $0.987^{+0.003}_{-0.003}$ & 96.76 & 82.59 & 52\\
$2.633$ & $1.488-1.496$ &$39.26-135.00$ & 5.9$\%$ &$0.995^{+0.004}_{-0.004}$ & 91.57 & 99.95 & 52\\
$2.145$ & $1.209-1.219$ &$39.26-135.00$ & 4.1$\%$ &$0.983^{+0.003}_{-0.003}$ & 99.3 & 94.56 & 52\\
$1.767$ & $0.992-1.003$ &$39.26-135.00$ & 5.4$\%$ &$0.988^{+0.004}_{-0.004}$ & 111.17 & 72.95 & 46\\
$1.303$ & $0.724-0.737$ &$39.26-135.00$ & 9.6$\%$ &$0.931^{+0.004}_{-0.003}$ & 112.96 & 87.30 & 45\\
$0.878$ & $0.479-0.495$ &$60.61-110.00$ & 7.4$\%$ &$1.077^{+0.006}_{0.006}$ & 28.2 & 29.31 & 29\\
$0.721$ & $0.385-0.403$ &$68.97-108.07$ & 6.1$\%$ &$1.025^{+0.007}_{-0.007}$ & 14.41 & 15.13 & 17\\
SONIK Total &  &  &  & & 849.35 & 1323.77 &451 \\
 \multicolumn{2}{c}{\textbf{(B) Barnard \textit{et al}.~\cite{Barnard_1964}}} & &&\\
$2.454-5.737$ & $1.39-3.27$ &$54.77-140.80$& 5$\%$ &$1.010^{+0.002}_{-0.002}$ & 1098.56 & 1996.24 & 646 \\
Total  & & & & & 1947.92 & 3320.01&1097 \\
\end{tabular}
\end{ruledtabular}
\label{chi_square}
\end{table*}

\begin{table}[tbp]
\renewcommand{\arraystretch}{1.5}
\caption{ The observed energies $\tilde{E}_{\text{x}}$ and reduced width amplitudes $\tilde{\gamma}$ obtained from the best \textit{R}-matrix fit with channel radius set at 4.2~fm. States in parentheses are introduced as background levels. The parameters in bold were treated as fit parameters and all others were held constant.}
\begin{ruledtabular}
\begin{tabular}{cccc}
\textit{J$^{\pi}$} & $l$ & $\tilde{E}_{\text{x}}$ (MeV) & \textit{$\tilde{\gamma}$}~~(MeV$^{1/2})$\\\hline
3/2$^{-}$ &1 & 0.000 & 0.931 \\
1/2$^{-}$& 1& 0.429 & 1.151\\
7/2$^{-}$&3 & $\mathbf{4.5639_{-0.0003}^{+0.0003}}$ & $\mathbf{0.924^{+0.003}_{-0.003}}$ \\
5/2$^{-}$&3 & 6.730 & 1.767 \\
1/2$^{+}$&0 & (14.000) & $\mathbf{1.683_{-0.004}^{+0.004}}$  \\
1/2$^{-}$&1 & (21.600) &  $\mathbf{-2.939_{-0.036}^{+0.038}}$ \\
3/2$^{+}$&2 & (12.000)  & $\mathbf{1.224_{-0.013}^{+0.013}}$ \\
5/2$^{+}$&2 & (12.000)  & $\mathbf{1.155_{-0.012}^{+0.012}}$ \\
3/2$^{-}$&1 & (21.600) & $\mathbf{-2.300_{-0.018}^{+0.019}}$  \\
\end{tabular}
\end{ruledtabular}
\label{R-matrix_param}
\end{table}


Using the MCMC-generated chain of $R$-matrix parameters and Eq.~(\ref{scatt_length}), the $s$-wave scattering length $a_0$ was calculated to be $33.10$ fm. The uncertainty from the MCMC analysis amounts to an uncertainty of $\pm0.13$ fm in the $s$-wave scattering length. Likewise, the MCMC results were used to calculate the effective
range function $K_L$ at $E$=0 and small positive energies.
The effective range is then obtained by numerical
differentiation.
The $s$-wave effective range $r_0$ was determined to be 1.009 fm. The uncertainty from the MCMC analysis yields an error bar of $\pm0.001$ fm.

\par
The sensitivity of the $s$-wave scattering parameters to the choice of excitation energy of the 1/2$^+$ level was studied at a fixed channel radius of 4.2 fm. A separate $R$-matrix fit in which  the excitation energy for the 1/2$^+$ state was allowed to vary was conducted using BRICK; this analysis is denoted SB$^+$ in the text hereafter. The excitation energy of the 1/2$^+$ background level resulting in the minimum total $\chi^2$ was found to be 9.22 MeV. However, the trend of the experimental 1/2$^+$ phase shift at higher energies determined by Spiger and Tombrello~\cite{Spiger_1967} is then not explained. The  $a_{0}$ and $r_{0}$ values obtained from the $SB^+$ analysis are 35.82$\pm$0.13~fm and 1.098$\pm$0.008 fm, respectively
  
The $s$-wave scattering parameters remain fairly constant with the choice of channel radius. The channel radius was varied between 3.8--4.6 fm keeping other parameters fixed, which resulted in $a_{0}$ and $r_{0}$
changing by 0.8 fm and 0.008 fm, respectively, from their values at a channel radius of 4.2 fm~\cite{SNPaneru}. 

The results quoted so far were obtained with the ANCs fixed to the same values that were used in Ref.~\cite{deBoer14}. We studied the implications of varying the ANCs for the inferred $a_{0}$ and $r_{0}$ parameters by considering the three sets of ANCs listed in Table I of Ref.~\cite{zhang_2020}. All three produce a change in the inferred $a_{0}$ and $r_{0}$ (relative to the SB analysis above) of $< 1$ fm and $< 0.01$ fm for  $a_{0}$ and $r_{0}$, respectively. Adopting the ANCs quoted from a recent measurement of Kiss \textit{et al}.~\cite{KISS2020} yields a change of +1.4 fm and +0.01 fm for $a_{0}$ and $r_{0}$.

We also studied the sensitivity of the scattering parameters to the choice of data sets and the energy range of the data set. The SONIK and Barnard \textit{et al.} data sets are represented by S and B, respectively. The sensitivity of the scattering parameters excluding the data above $E$[$^3$He]$<$4 MeV was  studied; the analyses using these energy-truncated data sets are represented by the superscript $(t)$.
The SB${}^+$ model is a superset of SB.
In the SB analysis, the background 1/2${}^+$ level is fixed at $E_x^{(1/2+)}=14$ MeV.
With SB${}^+$, we allow that parameter to vary between 2 and 20 MeV.
 
\begin{figure}
    \centering
    \includegraphics{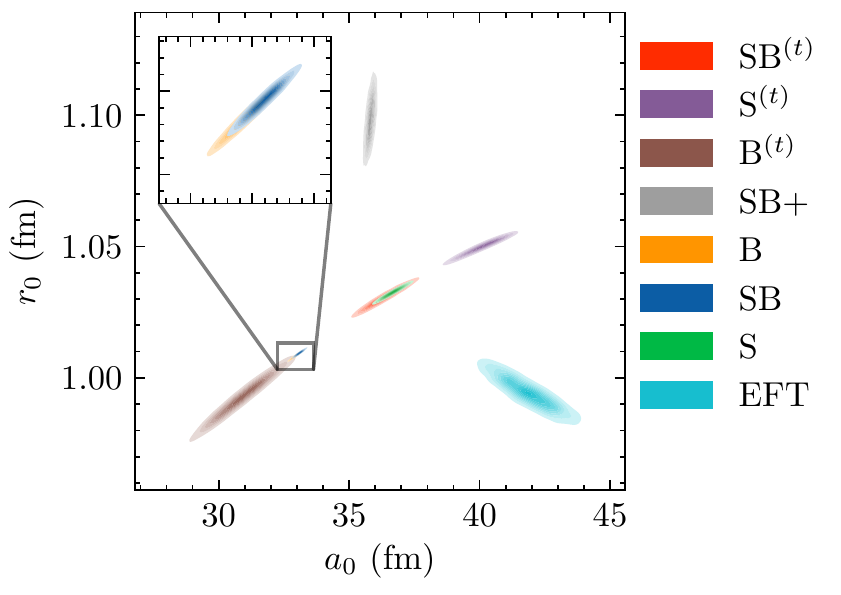}
    \caption{(Color online) Scattering length and effective range posterior probability distributions from each of the seven $R$-matrix analyses and the EFT analysis.}
    \label{fig:a0_r0_posterior}
\end{figure}

The results for $a_{0}$ and $r_{0}$ for several different $R$-matrix analyses and the EFT analysis described in this work are  depicted in Fig.~\ref{fig:a0_r0_posterior} and summarized in Table~\ref{tab:a0}.
Fig.~\ref{fig:a0_r0_posterior} reveals several interesting points.
First, of the seven different data models studied with $R$-matrix theory, six of them exhibit the same $a_0$-$r_0$ correlation.
Only the SB${}^+$ model breaks this consistency.
The additional freedom in the 1/2${}^+$ channel changes the correlation between $a_0$ and $r_0$ entirely.
Second, the EFT analysis displays a very different different correlation structure from all of the $R$-matrix analyses.
Finally, none of the truncated-data analyses encapsulate their associated complete-data analysis.
As more data is included, one expects a refinement of the previous result.
In this case, it is particularly striking that the inclusion of \textit{higher} energy data significantly changes the extracted low-energy scattering parameters.

\begin{table}[]
    \centering
    \begin{tabular}{r|c|c}
        Data Model & $a_0$ (fm) & $r_0$ (fm) \\
        \hline
        SB & $33.10_{-0.13}^{+0.13}$ & $1.009_{-0.001}^{+0.001}$ \\
        S & $36.67_{-0.36}^{+0.37}$ & $1.033_{-0.002}^{+0.002}$ \\
        B & $32.97_{-0.15}^{+0.16}$ & $1.009_{-0.001}^{+0.001}$ \\
        SB${}^{(t)}$ & $36.36_{-0.60}^{+0.61}$ & $1.031_{-0.004}^{+0.004}$ \\
        S${}^{(t)}$ & $40.10_{-0.74}^{+0.64}$ & $1.050_{-0.003}^{+0.003}$ \\
        B${}^{(t)}$ & $30.90_{-0.96}^{+0.95}$ & $0.993_{-0.008}^{+0.007}$ \\
        SB${}^{+}$ & $35.82_{-0.13}^{+0.13}$ & $1.098_{-0.008}^{+0.008}$ \\
        \hline
        Halo EFT & $41.89_{-0.89}^{+0.90}$ & $0.994_{-0.005}^{+0.006}$
    \end{tabular}
    \caption{Extracted scattering lengths and effective ranges--- with uncertainties--- from eight different data analyses. The first seven invoke different choices of $R$-matrix parameters and/or different data sets. The eighth is the analysis using Halo EFT that is described in the next section. }
    \label{tab:a0}
\end{table}
The $R$-matrix result for $a_{0}$, and $r_{0}$ is presented in the last line of Table~\ref{comp_scat_length}. It is obtained from a simultaneous $R$-matrix fit of all data from Barnard \textit{et al.} and the data of this work. This model, termed as SB, is our preferred model. The lower and upper limits in the $R$-matrix-extracted $a_{0}$ and $r_{0}$ values determined from the sensitivity studies listed in Table~\ref{tab:a0} 
and the variation of the ANCs are accounted for through an additional ``analysis error.'' This error is estimated as $^{+7.5}_{-3.0}$ fm, and $^{+0.096}_{-0.023}$ fm, respectively.  

\par
$s$-wave scattering parameters published in the literature are also presented in Table~\ref{comp_scat_length}. J. Dohet Eraly \textit{et al}.~\cite{DOHETERALY} used the chiral nucleon–nucleon interaction within the \textit{ab initio} no-core shell model with continuum (NCSMC) to calculate the $^3$He($^4$He, $\gamma$)$^7$Be astrophysical $S$-factor and deduced the $s$-wave scattering length. The scattering parameters have also been calculated using a microscopic cluster model~\cite{Kamouni2007}. The scattering parameters for the $^3$He+$^4$He system have been extracted from a Bayesian analysis of the capture data below 2 MeV that used Halo EFT~\cite{zhang_2020}. Premarathna and  Rupak also performed a Bayesian analysis of the capture data and the phase shifts from Boykin \textit{et al}~\cite{Boykin_1972} to infer the scattering parameters~\cite{Rupak_2020}.  The $R$-matrix fit to the SONIK and Barnard data yields an $s$-wave scattering length and $s$-wave effective range in fair agreement (within 1.5$\sigma$) with all but one of these values previously published in the literature, provided the dispersion of $a_{0}$, and $r_{0}$ values with respect to the choice and energy range of data sets included in the analysis is considered. The exception is the NSCMC calculation of Ref.~\cite{DOHETERALY} which obtained a much smaller scattering length than was found in any of the data analyses or in the microscopic cluster model. 

\begin{table*}[tbp]
\renewcommand{\arraystretch}{1.5}
\begin{center}
\caption{$s$-wave scattering parameters for the $^3$He+$^4$He system.}
\begin{tabular}{llll}\hline\hline
$a_{0}$ (fm) & $r_{0}$ (fm)& Method & Reference\\\hline
7.7 & - & NCSMC &~\cite{DOHETERALY}\\
41.06 &1.01 & Microscopic &~\cite{Kamouni2007}\\
 & & Cluster Model &\\
40$^{+5}_{-6}$& 1.09$^{+0.09}_{-0.1}$ & Halo EFT &~\cite{Rupak_2020}\\
$50^{+7}_{-6}$ & 0.97$\pm$ 0.03 &Halo EFT &~\cite{zhang_2020}\\
$42_{-1}^{+1}$ & $0.994^{+0.006}_{-0.005}$ &${\rm Halo~EFT}$ & This Work\\
$33.10 \pm 0.13 (\text{stat}) ^{+7.5}_{-3} (\text{analysis})$ & 
$1.009 \pm 0.001(\text{stat}) ^{+0.096}_{-0.023} (\text{analysis})$ &$R$-matrix &This Work\\\hline\hline
\end{tabular}
\label{comp_scat_length}
\end{center}
\end{table*}

\begin{figure}
    \centering
    \includegraphics[width=\columnwidth]{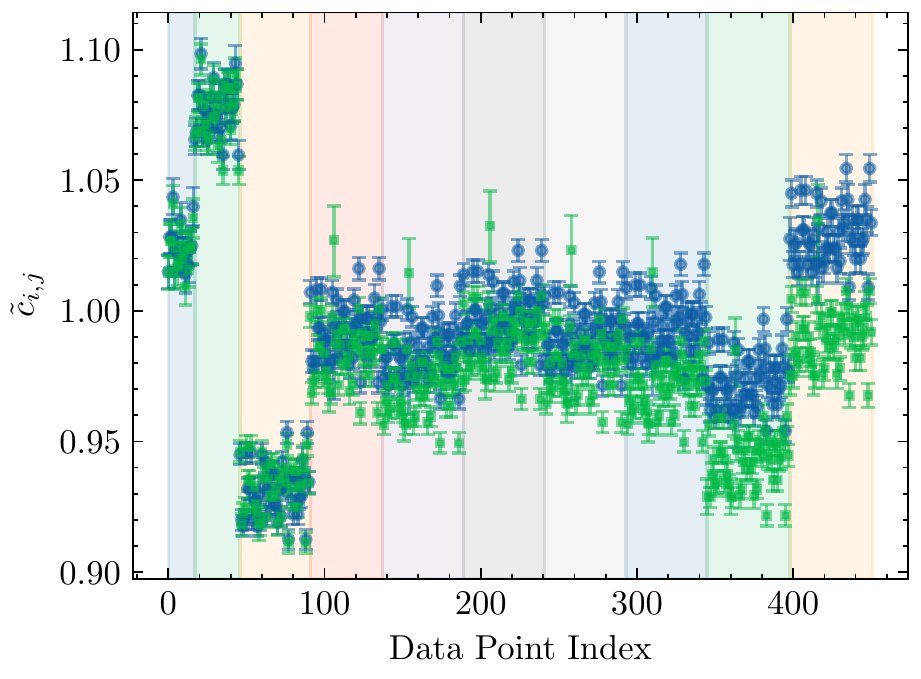}
    \caption{Normalization factor results at each SONIK data point are shown as a product of the three different systematic effects described in subsection~\ref{bayesian_analysis}. The results are shown in increasing energy (from left to right) as a function of data point index. Each vertical, shaded region corresponds to a different energy bin. $R$-matrix results are shown as blue circles with error bars. EFT results are shown as green squares with error bars.} \label{fig:c_tilde_summary}
\end{figure}

The results for the different normalization factors applied to the SONIK data in the SB analysis are summarized in Fig.~\ref{fig:c_tilde_summary}.
The results from the $R$-matrix (blue) and EFT (green) analyses of the SB data model are in good agreement at low energies (small values of the data point index).
While significant effort was put into accounting for detector-specific systematic effects, the overall result, $\tilde{c}_{i,j}$ clearly shows that the energy-dependent systematics dominate in both analyses.
While the EFT normalization factors tend toward lower values than the $R$-matrix ones as the energy increases (larger data point indices), the ``bunching'' of $\tilde{c}_{i,j}$ with respect to the energy bins --- indicated by vertical, colored bands --- is consistent between both theories.
These results are only for the SONIK data.
The Barnard data set lacked the necessary uncertainty information to apply such a detailed treatment of its systematic error.

\subsection{Results from Halo EFT}\label{res:haloeft}

Table~\ref{comp_scat_length} also includes the $s$-wave scattering parameters from a NNLO Halo EFT analysis of the same data set as was used in the $R$-matrix analysis. This fit reproduces the SONIK data well, especially for c.m. energies below 2 MeV. The band of cross-sections generated from the posterior samples obtained from sampling the EFT likelihood, Eq.~(\ref{eq:likelihood2}), for each SONIK energy bin and Barnard angular bin are shown respectively in Figures~\ref{Fits_SONIK_1}, \ref{Fits_SONIK_2}, \ref{Fits_SONIK_3}, and \ref{Fits_barnard}. The $\chi^2_{\rm EFT}$ values obtained from this analysis for each energy bin are provided in Table \ref{chi_square}. Note that $\chi^2_{\rm EFT}$ is generically less than the standard $\chi^2$, because it includes a theory-error piece of the covariance matrix (see Eq.~\ref{eq:chi-square2}). For comparison the total $\chi^2$ for the EFT fit to the SONIK data is 2165, as compared to $\chi^2_{\rm EFT}=1324$. 
Most of the difference between $\chi^2_{\rm EFT}$ and the standard $\chi^2$ accumulates above $E~[{}^3{\rm He}]=1.767$ MeV.

The EFT analysis also accumulates large $\chi^2_{\rm EFT}$ in the higher SONIK energy bins (especially at backward angles) 
and for the portion of the Barnard data in and beyond the ${\frac{7}{2}}^-$ resonance. For the latter data set the largest $\chi^2$ contribution comes from the $140.80^\circ$ bin. The large value of $\chi^2$ around the ${\frac{7}{2}}^-$ resonance, especially at backward angles, suggests that the approach of Ref.~\cite{Poudel_2022} does not adequately describe that  resonance. (In fact, the analysis of Ref.~\cite{Poudel_2022} did not include the data in the highest SONIK energy bin, because the EFT is not tailored to that region.). 
If we choose to sample the width of the resonance, $\Gamma^{{\frac{7}{2}}^-}$, in the EFT calculation we obtain $151$ keV, not the $159$ keV used to produce the results presented here. The sizable $\chi^2_{\rm EFT}$ in the EFT fit in the vicinity of the resonance could likely be improved by better parameter estimation or a better model of the resonance. 
The inference of the $s$-wave parameters is also surprisingly sensitive to the description of the ${\frac{5}{2}}^-$ phase shift. If the EFT analysis is performed with $\Gamma^{{\frac{5}{2}}^-}=1.9$ MeV a larger $a_0$ (and a smaller $r_0$) are obtained. The effect of such changes in the $f$-wave phase shifts on the $s$-wave ERPs is {\it not} quantified in the uncertainties we provide here. 
The median values of the posterior samples for effective range parameters (ERPs) from this analysis are as follows: $a_0 = 42^{+1}_{-1}~{\rm fm}, r_0 = 0.994^{+0.006}_{-0.005}~{\rm fm}, P_{1^+}=1.681^{+0.005}_{-0.005}~{\rm fm}, P_{1^-}=1.810^{+0.009}_{-0.010}~{\rm fm}$. This analysis finds that $a_0$ and $r_0$ are anti-correlated, with a correlation coefficient of -0.92.\par
Returning to Table~\ref{comp_scat_length}, the central value of $a_0$ from the Halo EFT analysis of SONIK and Barnard data is very different from that predicted in the NCSMC ({\it ab initio}) calculation of Ref.~\cite{DOHETERALY} while consistent with the result of the microscopic calculation of Ref.~\cite{Kamouni2007}.
Compared to the other Halo EFT analyses of data listed in Table~\ref{comp_scat_length}, the 68\% interval for $a_0$ found in this analysis completely falls within the distribution of $a_0$ obtained in Ref.~\cite{Rupak_2020}. But our 68\% $a_0$ interval  does not overlap the one obtained from ${}^3$He($\alpha$,$\gamma$) data in Ref.~\cite{zhang_2020}. Meanwhile, the $r_0$ values are consistent between the microscopic prediction of Ref.~\cite{Kamouni2007} and the Halo EFT analyses of Refs.~\cite{Rupak_2020, zhang_2020}. 
Also, the $r_0$ value reported here is consistent with the natural scale of $\frac{1}{\Lambda}$ assigned by EFT. 

\section{CONCLUSIONS}
\label{conclusion}

    \begin{figure}
        \centering
        \includegraphics{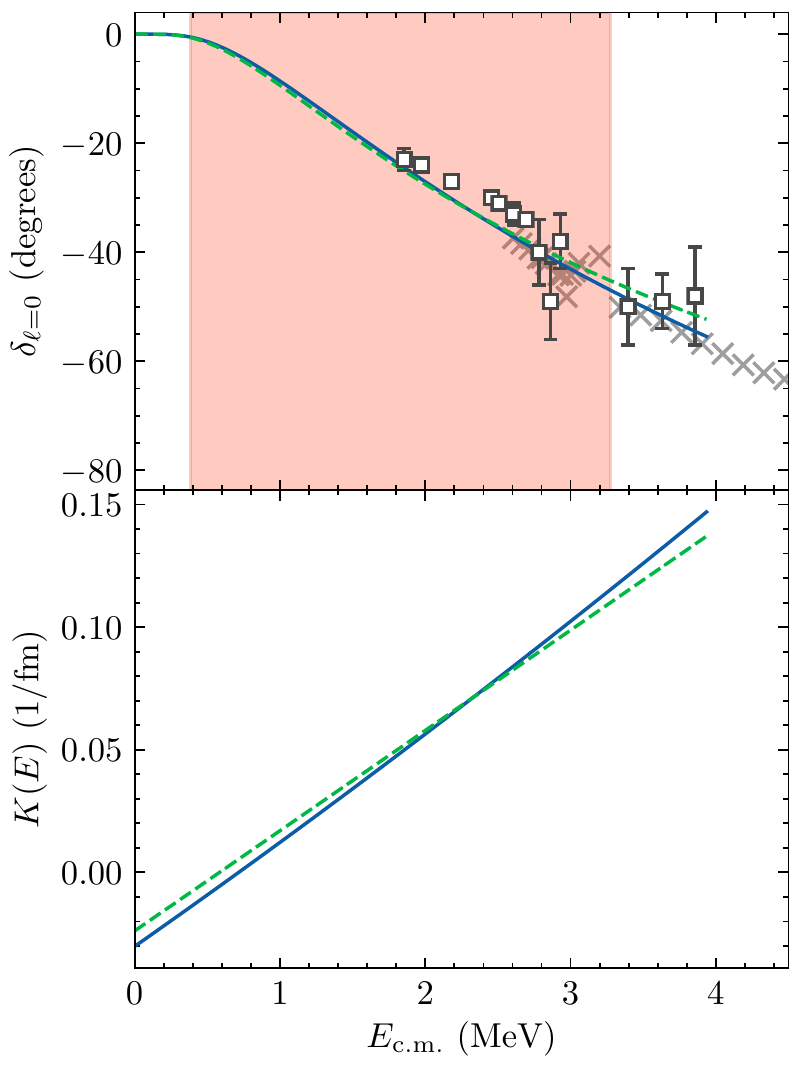}
        \caption{Top: The scattering phase shifts for $\ell=0$ are shown in comparison to the analyses in Ref.~\cite{Boykin_1972} and Ref.~\cite{Spiger_1967}. The solid, blue line represents the median calculated in the SB $R$-matrix analysis. The dashed, green line represents the median calculated in the EFT analysis. The red, shaded region indicates the energy range over which the SONIK measurements were carried out. White squares with error bars and grey x's indicate the analyses of \cite{Boykin_1972} and \cite{Spiger_1967}, respectively. Bottom: The effective range function, $K(E)$, is plotted as a function of the center of mass energy.}
        \label{fig:swave_phase_shifts}
    \end{figure}

The elastic scattering reaction $^4$He($^3$He,$^3$He)$^4$He was measured at 9 different energies from $E_{\text{c.m.}}=0.38 -3.13 $ MeV. This data set includes the first measurement of  elastic scattering in the $^3$He+$^4$He system 
below $E_{\text{c.m.}}=500$ keV. The angular range covered is
$30^{\circ}<\theta_{\text{c.m.}}<139^{\circ}$, a wider range than in previous measurements.
This elastic scattering measurement of $^4$He($^3$He,$^3$He)$^4$He is the first scientific measurement made using SONIK. Its success validates the use of SONIK for charged particle scattering measurements. 
  The resulting data is presented in this paper, together with detailed error estimates which are lacking in previous measurements of ${}^3$He + ${}^4$He elastic scattering. They are consistent with previous experimental measurements and have better precision. 
 \par 
The extraction of $s$-wave effective-range
parameters for the $^3$He+$^4$He system from these data was carried out using both an $R$-matrix and a Halo EFT analysis. We used the 
Bayesian $R$-matrix Inference Code Kit {\sc{BRICK}}~\cite{Odell:2021nmp} to calibrate the $R$-matrix model against the data of this work and the elastic scattering data of Ref.~\cite{Barnard_1964}.
The $R$-matrix parameter posteriors were then employed to calculate the $s$-wave scattering length and effective range, fully propagating the model uncertainties to these extracted quantities. This yields $a_0=33.10 \pm 0.13 ~(\text{stat}) ^{+7.5}_{-3} (\text{analysis})$ fm.
The same combined SONIK + Barnard scattering data set was analyzed using Halo EFT at NNLO---also with full uncertainty quantification. The result $a_0=42 \pm 1$ fm is obtained in that approach.
The two analyses thus yield discrepant values for $a_0$, with a concomitant discrepancy in their results for $r_0$. The $s$-wave scattering length from the $R$-matrix analysis is in fair agreement with the prediction of the microscopic cluster model~\cite{Kamouni2007}, and previous inferences from data using Halo EFT~\cite{Rupak_2020, zhang_2020}.
\par
The discrepancies in the inferred $a_0$ and $r_0$ values naturally suggest an examination of the $s$-wave phase shifts. The $s$-wave phase shifts from the $R$-matrix and Halo EFT analyses are compared in the upper panel of Fig.~\ref{fig:swave_phase_shifts}. The two analyses agree with each other over most of the energy range of the SONIK data. At the energies of the Boykin et al.~\cite{Boykin_1972} phase shifts both analyses yield lower phase shifts than were reported in that work. The phase shifts inferred using $R$-matrix and Halo EFT begin to diverge a little at the upper end of the SONIK energy range. This is related to the fact that, at slightly higher energies, the $R$-matrix analysis describes the phase shifts from Ref.~\cite{Spiger_1967} much better than the Halo EFT result does. These differences in the phase shift at high energy then affect the behavior at low energy, as is evident from the $s$-wave effective range function shown in the lower panel of Fig.~\ref{fig:swave_phase_shifts}. 

The EFT of Ref.~\cite{Poudel_2022} that was constructed to describe the elastic scattering reaction measured in this experiment breaks down at backward angles for the higher-energy data bins. In Ref.~\cite{Poudel_2022} Poudel and Phillips attempted to mitigate this via phenomenological inclusion of the $7/2-$ resonance. Even though that was done here too, the $\chi^2$ of the EFT fit to data is strikingly large for $E_{\rm c.m.} > 2$ MeV, in spite of the addition of a theory component of the errors in the $\chi^2$. Future work to build a better EFT description through the $7/2-$ resonance (cf. Ref.~\cite{Higa:2022mlt} for the case of the ${}^7$Be-proton system) is needed. 
The quality of the $R$-matrix fit also deteriorates in the final energy bin of the data set described here, and the value of $a_0$ inferred is higher if only data below $E_{\rm c.m.}=2.2$ MeV is used, see Fig.~\ref{fig:a0_r0_posterior}. Future studies should address whether Halo EFT and $R$-matrix agree if only low-energy data is used. 

A better measurement of elastic ${}^3$He-${}^4$He scattering in the vicinity of the 7/2$^-$ resonance, as well as an accurate determination of the position of this resonance and its $\alpha$-width may also help resolve the discrepancy between the Halo EFT and $R$-matrix analyses of the SONIK data. The inference of $s$-wave parameters is surprisingly sensitive to the description of the 5/2$^-$ phase shift and the width of 5/2$^-$ level. This issue could be explored further by a similar analysis with the addition of scattering data above the proton separation energy in $^7$Be.
\par
 Ultimately, smaller uncertainties in the scattering parameters will reduce the overall uncertainty in $S_{34}(0)$, just as  smaller uncertainties in the $s$-wave scattering lengths for $^7$Be+$p$ from Ref.~\cite{Paneru} led to reduced uncertainty in $S_{17}(0)$ presented in Ref.~\cite{Higa2020}.
The data from this measurement should be used in global $R$-matrix and Halo EFT analyses of $^4$He($^3$He,$\gamma$)$^7$Be data, in order to resolve disagreements between previous analyses
regarding $S_{34}(0)$ and reduce the extrapolation error therein.

\section*{Acknowledgements}

We are grateful to R. J. deBoer for his expertise on ${}^3{\rm He}+{}^4{\rm He}$ scattering and reactions and assistance with {\sc{Azure2}}. We are indebted to Keerthi Jayamana and the TRIUMF OLIS staff for providing the intense and pure $^3$He beam, and we are grateful to the TRIUMF operations staff whose hard work made the experiment possible. This work was supported in part by the U.S. Department of Energy under Grants No. DE-NA0003883, and No. DE-FG02-93ER40789. The Canadian authors are supported by the Natural Sciences and Engineering Research Council of Canada under grants: SAPPJ-2016-00029 and SAPPJ-2019-00039. TRIUMF receives federal funding through a contribution agreement with the National Research Council of Canada. This research was supported in part by the Notre Dame Center for Research Computing.

\bibliography{PRCreferences}
\end{document}